\documentclass[aps,pre,notitlepage,reprint,onecolumn,showpacs,superscriptaddress]{revtex4-1}
\usepackage{epsfig}
\usepackage{natbib}
\usepackage{ulem}
\usepackage{xspace}
\usepackage{xcolor}
\usepackage{graphicx}
\usepackage{amssymb}
\usepackage{cancel}
\usepackage{lineno}
\usepackage{amsmath}
\usepackage{stackrel}
\usepackage{appendix}
\usepackage{url}
 \usepackage{enumitem}
 \usepackage{hyperref}
\usepackage{xcolor}
\usepackage{array}
\usepackage{booktabs}
\usepackage{multirow}

\hypersetup{
	colorlinks=true,
	linkcolor=blue,
	citecolor=blue,
	urlcolor=blue
}

\begin{document}
\title{Solutions of first passage times problems: a bi-scaling approach}

\author{Talia Baravi}
\affiliation{Department of Physics, Institute of Nanotechnology and Advanced Materials, Bar-Ilan University, Ramat Gan 52900, Israel}
  \author{David A. Kessler}
           \affiliation{Department of Physics, Br-Ilan University, Ramat Gan 52900, Israel}
\author{Eli Barkai}
\affiliation{Department of Physics, Institute of Nanotechnology and Advanced Materials, Bar-Ilan University, Ramat Gan 52900, Israel}  

\begin{abstract}
 We study the first-passage time (FPT) problem for widespread recurrent processes in confined though large systems and present a comprehensive framework for characterizing the FPT distribution over many time scales. We find that the FPT statistics can be described by two scaling functions: one corresponds to the solution for an infinite system, and the other describes a scaling that depends on system size. We find a universal scaling relationship for the FPT moments $\langle t^q \rangle$ with respect to the domain size and the source-target distance. This scaling exhibits a transition at $q_c=\theta$, where $\theta$ is the persistence exponent. For low-order moments with $q<q_c$, convergence occurs towards the moments of an infinite system. In contrast, the high-order moments, $q>q_c$, can be derived from an infinite density function. The presented uniform approximation, connecting the two scaling functions, provides a description of the first-passage time statistics across all time scales. We extend the results to include diffusion in a confining potential in the high-temperature limit, where the potential strength takes the place of the system's size as the relevant scale. This study has been applied to various mediums, including a particle in a box, two-dimensional wedge, fractal geometries, non-Markovian processes and the non-equilibrium process of resetting.

\end{abstract}
\maketitle

\section{Introduction}
The concept of first passage time (FPT) in bounded domains is crucial in various stochastic processes \cite{redner2001guide,TargetSearch2024,bray2013persistence,metzler2004restaurant,meyer2011universality,benichou2014first,basu2018active,boyer2004modeling,jolakoski2023first,lanoiselee2018diffusion,shlesinger2006search,guerin2016mean,grebenkov2023boundary,scher2023escape,padash2022asymmetric,dubkov2023enhancement}, such as chemical reactions \cite{schuss2007narrow, mirny2008cell, benichou2008optimizing}, animal behavior \cite{giuggioli2014consequences, benichou2005optimal, benichou2006two}, and disease spreading \cite{sugaya2018analysis, gallos2007scaling}.
Condamin \textit{et al.}~\cite{condamin2007first}, building on the O'Shaughnessy-Procaccia model for diffusion on fractals~\cite{o1985diffusion,o1985analytical}, proposed a universal FPT distribution for diffusion in realistic confining geometries. Their theory, based on a mono-scaling ansatz, uses dimensional analysis to identify a time scale \(\tau_L\) that diverges with system size \(L\). The FPT distribution could be expressed as a mono-scaling function \(\tilde{N} h(t/\tau_L)\), where \(\tilde{N}\) is a normalization factor. However, as later pointed out by Meyer \textit{et al.} \cite{meyer2011universality}, the function is, in fact, non-normalizable, indicating a limitation in the original approach. 
Acknowledging these shortcomings, we extend this framework by introducing a bi-scaling theory that better captures the multi-timescale nature of FPT, particularly in recurrent (compact) stochastic processes. Our approach, which employs two timescales determined by the initial distance from the target and the confinement length, provides a more accurate understanding of FPT in bounded systems, addressing the inadequacies of the traditional mono-scaling approach.

\par Physically, once we identify the pair of time scales in the system, a general bi-scaling theory suggests the existence of a scaling function \(\tilde{N} h(t/\tau_L, t/\tau_0)\), where \(\tau_0\) depends on the initial distance between the particle and the target. Remarkably, in the accompanying letter, we demonstrated that for compact search in diverse systems, the solution factorizes. This means that the first passage time PDF can be expressed as a product of two functions, $f_{\infty}(\cdot)$ and \(\mathcal{I}(\cdot)\), which are defined below. This factorization is a significant simplification and has profound implications for the generic features of observables and the foundational aspects of first passage time theory. To solve compact first passage time problems in the scaling limit, it is essential to determine or measure these two functions, each carrying special significance.
\par Another result of the bi-scaling theory is the scaling of the FPT moments. As mentioned, the theory is built upon two distinct time scales, which are determined by two corresponding length scales. Specifically, let $t$ be the FPT, $r_0$ is the initial distance from the target, and $L$ is the measure of the system size or the domain explored by the particles (see below). We showed that the FPT statistics exhibits bi-scaling \cite{castiglione1999strong}, implying 
\begin{equation}\label{eq:intro1}
    \langle t^q\rangle \propto r_0^{q\mu(q)}L^{q\nu(q)} \, ,
\end{equation}
where
\begin{align}\label{eq:intro2}
    q\nu(q) &=\begin{cases}
                     0 , &  q<\theta\\
                    d_w(q-\theta), &  q>\theta\\
                    \end{cases} \, , &
    q\mu(q) &=\begin{cases}
                     d_wq , &  q<\theta\\
                    d_w\theta, &  q>\theta\\
                    \end{cases} \, .
\end{align}
Here $q\geq 0$, $d_w$ is the walk dimension and $\theta$ is the persistence exponent. This indicates that $d_w$ and $\theta$, which are well known exponents modeling diverse 
systems \cite{havlin1987diffusion}, are crucial in the determination of the behavior of the moments. Further, we witness a sharp non-analytical transition, at $q=\theta$, in the spectrum of exponents describing this problem.
\par As mentioned, the problem involves two distinct scaling functions. The first, \( f_{\infty}(\cdot) \), is the solution for an infinite system, a well-known problem. This solution does not depend on the confinement length \( L \). The second, $\mathcal{I}(\cdot)$, is a non-normalizable function that describes intermediate to long times. While the calculation of \(f_{\infty}(\cdot)\) has been extensively studied in numerous works, much less is known about \(\mathcal{I}(\cdot)\). We showed that this second scaling solution does not depend on the initial condition. Non-normalized densities, \(\mathcal{I}(\cdot)\), are also observed in other cases \cite{afek2023colloquium,aghion2019non,korabel2009pesin,barkai2021transitions}. However, this concept, which originates from infinite ergodic theory, has not been prominently highlighted in the context of the FPT problem.

Our work focuses on calculating this second scaling function for various systems, including simple example of a particle in a box, wedges, semi-Markovian dynamics, fractal geometries, resetting process, run-and-tumble particle and diffusion in a force field. We will show how the two function $f_{\infty}(\cdot)$ and $\mathcal{I}(\cdot)$ match for compact search and how to use these functions within a uniform approximation for the calculation of observables like moments. By employing analytical techniques such as the WKB approximation, eigenfunction expansions, Mellin transform and the summation of infinite series of exponential decays, we derive scaling functions that accurately describe the FPT distributions across different timescales. All along this work we back the theoretical results with numerical simulations. In these problems, we can treat the limiting distributions as $L$ approaches infinity while keeping $t$ fixed, which yields nothing new to the solution for an infinite system. By taking a second limit where both $L$ and $t$ are large, we reach a limit where the solutions become non-normalized. Rather than dismissing these functions as non-useful, we show they effectively describe FPTs in compact domains. Hence, a main goal in calculating the FPT PDF is to evaluate these functions. We offer a versatile theory that enhances both theoretical understanding and practical applications, providing tools to understand diffusion processes in diverse scientific and engineering fields.

\par The structure of this paper is outlined as follows. In Section 2, we derive the non-normalized scaling function for the FPT in the straightforward case of a particle confined to a one-dimensional (1D) box, This model already shows the relevant behaviors that are also found in the other models. In Section 3 we present a bi-scaling theory for the FPT in confined geometries, which include the bi-scaling property of the probability density and introduces the different scaling functions. Sections 4 to 6 focus on specific results obtained for a two-dimensional (2D) wedge, non-Markovian dynamics described by continuous-time random walks (CTRW), and fractal geometries, respectively. In Sections 7-8, we delve into the presence of the scaling functions in scenarios involving weak confining fields. Section 9 addresses the solution to the FPT problem under stochastic resetting, while Section 10 examines the dynamics of a run-and-tumble particle with resetting. Finally, in Section 11, we expand upon the uniform approximation method to the FPT PDF and its validity.

\section{ Particle in a one-dimensional box \label{sec:1D-diff}} 
To illustrate our approach, we begin with a straightforward example involving a one-dimensional Brownian particle confined within a finite interval $x \in [0,L]$. The particle is initially located at position $x_0>0$ at time $t=0$. Our objective is to determine the PDF for the FPT of the particle reaching the origin at $x=0$. The case of an infinitely large interval ($L\rightarrow \infty$) has been previously addressed by Schrödinger \cite{schrodinger1915theorie, redner2001guide}, resulting in the PDF given by 
\begin{equation}\label{eq:freesol}
    \eta(t)_\mathrm{free}=\frac{x_0}{\sqrt{4\pi D}} t^{-3/2}\exp\left(-\frac{x_0^2}{4D t}\right) \, .
\end{equation}
Here the subscript free implies a system independent of reflecting  walls. The $q$-moment of the FPT in semi-infinite domain $x\in[0,\infty)$ in Eq.\,(\ref{eq:freesol}), defined as $\langle t^q\rangle_\mathrm{free}=\int_0^{\infty}t^q\eta(t)_{free}dt$, is given by
\begin{equation}\label{asymp0}
    \langle t^q\rangle_{\mathrm{free}} =\begin{cases}
                     (\frac{x_0^2}{4 D})^q\frac{\Gamma(1/2-q)}{\sqrt{\pi}}
 , &  q<\frac{1}{2}\\
                     \infty , &  q>\frac{1}{2}\\
                    \end{cases} \, ,
\end{equation}
where we restrict the discussion to $q\geq 0$. From Eq. (\ref{asymp0}) we see that the mean first-passage time (MFPT), i.e. the case $q=1$, diverges. In the case of a finite system with size $L$, Eq. (\ref{eq:freesol}) remains valid only for short time scales $t \ll L^2/D$. However, for longer times in the range of $t \sim L^2/D$, the validity of Eq. (\ref{eq:freesol}) breaks down due to the effect of the boundary. 

To account for the boundary effects in a finite system, we now introduce two methods to calculate the scaling function that captures this behavior.
The first method involves an expansion of the solution in Laplace space. In the second method, we use the inverse Mellin transform of the FPT moments to find the scaling function. The first method provides a straightforward and rigorous calculation, while the second method, although subtly different, offers a clearer physical picture of the role of the scaling function. The infinite density function can also be obtained by a direct calculation of the eigenfunction expansion as seen in appendix \ref{appendix:1d}. 
\par For the one-dimensional (1D) Brownian particle, the governing diffusion equation is
\begin{equation}\label{eq:FPeq1}
    \partial_t P(x,t)=D\partial_x^2 P(x,t) \, .
\end{equation}
The exact PDF for the FPT of a particle is determined by the boundary conditions applied to the system. We impose an absorbing boundary at the origin and a reflective boundary at $x=L$. The PDF $\eta(t)$ describes the probability of the time the particle reaches the absorbing boundary at $x=0$ for the first time. The exact expression for the Laplace transform of the PDF, defined by $\hat{\eta}(s)=\int_0^\infty dte^{-st}\eta(t)$, is given by \cite{redner2001guide}
\begin{equation}\label{eta10}
    \hat{\eta}(s)= \cosh\left(\sqrt{\frac{s}{D}}x_0\right)-\sinh\left(\sqrt{\frac{s}{D}}x_0\right)\tanh\left(\sqrt{\frac{s}{D}}L\right) \, .
\end{equation}
To perform the Inverse Laplace transform, we note that the only nonanalyticity of $\hat{\eta}(\cdot)$ lie on the negative real axis, due to zeros of the $\tanh$.  By Cauchy's theorem, then
\begin{equation}
\eta(t) = \sum_{n=0}^\infty  \frac{D\sin\left(\left(n+\frac{1}{2}\right)\frac{\pi x_0}{L}\right)(2n+1)\pi}{L^2} e^{-(n+1/2)^2\pi^2 Dt /L^2}
\end{equation}
We are interested in the limit of large, but finite $L$, fixed $x_0$.  This sum can be approximated by an integral as long as $t\ll L^2$, in which the summand is not a smooth function of $n$.  Doing this yields
\begin{equation}
\eta(t) \approx \frac{x_0}{\sqrt{4\pi D t^3}} e^{-x_0^2/4Dt}
\end{equation}
which is independent of $L$, and is the $L\to \infty$ limit of the problem, yields the free-particle solution given in Eq.\,(\ref{eq:freesol}). However, it is clear that this is wrong for large $t$, where the sum is dominated by the first term and decays exponentially in $t$, as opposed to the $t^{-3/2}$ asymptotics of the $L\to\infty$ limit.  For $t\sim L^2/D$,
however, we can simplify the sum since the argument of the {\it sin} is always small as long as $x_0/L \ll 1$. This yields the large-$t$ regime result
 \begin{align}\label{eq:etasum1d0}
    \eta(t)\simeq 2\frac{x_0}{L^3}\sum_{n=0}^\infty \pi^2(n+\frac{1}{2})^2D e^{-\pi^2(n+\frac{1}{2})^2Dt/L^2}\, \\ \label{eq:etasum1d}
    = -2\frac{x_0}{L}\partial_t\sum_{n=0}^\infty e^{-\pi^2(n+\frac{1}{2})^2Dt/L^2}\, 
\end{align}

The FPT statistics involve two distinct timescales: \(\tau_0 = x_0^2/D\), representing the characteristic time associated with the initial position, and \(\tau_L = L^2/D\), representing the characteristic time associated with the system size. It is useful to define a random variable $\tau\equiv t/\tau_L$ as the rescaled time, so in the limit of $L\rightarrow\infty$ and $t\rightarrow \infty$, this new variable remain finite. In this limit, we define the scaling function $\mathcal{I}(\tau)$ as
\begin{equation}\label{eq:infden1d}
\mathcal{I}(\tau)\equiv\lim_{t,\tau_L \rightarrow \infty} \tau_L^{3/2}\tau_0^{-1/2}\, \eta(t)\, .
\end{equation}
Summing the series in Eq. (\ref{eq:etasum1d}), we obtain
\begin{equation}\label{eq:inf1first}
 \boxed{ \mathcal{I}(\tau)= -\partial_{\tau }\vartheta_2(e^{-\pi^2 \tau}) \, ,}
\end{equation}
where $\vartheta_a(.)$ is the Jacobi elliptic theta function, with $a=2$. We will now discuss in details the meaning of $\mathcal{I}(\tau)$. 
\par As we will soon demonstrate, the function $\mathcal{I}(\tau)$ can be used to find the moments $\langle t^q \rangle$ with $q\geq 1/2$, however, this function is not normalized as $\int_0^\infty d\tau \mathcal{I}(\tau)\rightarrow \infty$. This is related to asymptotic properties of the Jacobi elliptic theta function for small $\tau$, which we will discuss shortly. Hence the scaling function $\mathcal{I}(\tau)$ is not a probability density but an infinite density. The term infinite simply implies that $\mathcal{I}(\tau)$ is not normalized, and as mentioned in the introduction, this function is employed in other fields. Despite this, $\mathcal{I}(\tau)$ is still related to the normalized probability density as given in Eq. (\ref{eq:infden1d}), since both functions provide the moments of the process, though the former only for $q>1/2$ as we will shortly explain. 
\par Using the asymptotic of the Jacobi elliptic theta function  $\vartheta_2(\tau)$, Eq. (\ref{eq:inf1first}) provides two distinct time dependencies in the limits of $\tau\ll 1$ and $\tau \gg 1$,
\begin{equation}\label{asymp1}
    \vartheta_2(e^{-\pi^2 \tau}) \sim \begin{cases}
 (\pi \tau)^{-\frac{1}{2}}  
 , &  \tau\ll 1\\
            2 e^{-\frac{\pi^2}{4} \tau}          , &  \tau \gg 1
                    \end{cases} \, ,
\end{equation}
thus, the asymptotics of the infinite density function are given by
\begin{equation}\label{asymp2}
    \mathcal{I}(\tau) \sim \begin{cases}
  \tau^{-3/2}  /\sqrt{2 \pi}
 , &  \tau\ll 1\\
            \frac{\pi^2}{2} e^{-\frac{\pi^2}{4} \tau}          , &  \tau \gg 1
                    \end{cases} \, .
\end{equation}
From the $\tau \ll 1$ limit, $\mathcal{I}(\tau)\sim \tau^{-3/2}$, it became clear that this function is not normalizable as mentioned. Using Eq. (\ref{eq:infden1d}), we find that the PDF for the FPT is given approximately by $\eta(t)\sim \tau_L^{-3/2}\mathcal{I}(\tau)$, which allows us to use Eq. (\ref{asymp2}) to study the asymptotics of $\eta(t)$. In the limit $\tau_0 \ll t \ll \tau_L$,  
\begin{equation}\label{}
    \eta(t)\sim \frac{x_0}{\sqrt{2D\pi}}t^{-3/2} , 
\end{equation}
which is equivalent the large $t$ limit of the PDF for the infinite system given in Eq. (\ref{eq:freesol}). In the limit of long times, $t\gg \tau_L$, 
\begin{equation}\label{}
    \eta(t)\sim \frac{\pi^2}{2}\sqrt{\tau_0}t^{-3/2} e^{-\frac{\pi^2}{4} t/\tau_L} ,
\end{equation}
which can be obtained using the smallest eigen-value of the diffusion operator. Indeed some approximations of the first passage time problem use the lowest eigen-value method,  which here means an exponential decay of the PDF, however this describes the rarest events, while the infinite density approach covers a by far broader scale of times.
\par In figure \ref{fig:1dbox} we observe the converges of the rescaled solution $\tau_L^{3/2}\tau_0^{1/2}\eta(t)$ toward the infinite density function. We see that indeed $\mathcal{I}(\tau)$ describes successfully the dynamics of the system from the intermediate time power-law regime to the long-time exponential decay associated with rare events. 
\par To conclude, we see that the limit of large systems can be considered by two routes. The first is to take $L\rightarrow \infty$, and fix the first passage time, and hence the famous Schrödinger law given in Eq. (\ref{eq:freesol}) holds. The second is to consider both $t$ and $L$ large, and then we get the infinite density $\mathcal{I}(\tau)$. It is rewarding that we were able to find a non-trivial limit for such a simple problem, as diffusion in a box, but more importantly, that these limits exist for a variety of systems as shown in the following sections.

\subsection{The FPT moments}
\begin{figure}[h]
    \centering
    \includegraphics[width=1\textwidth]{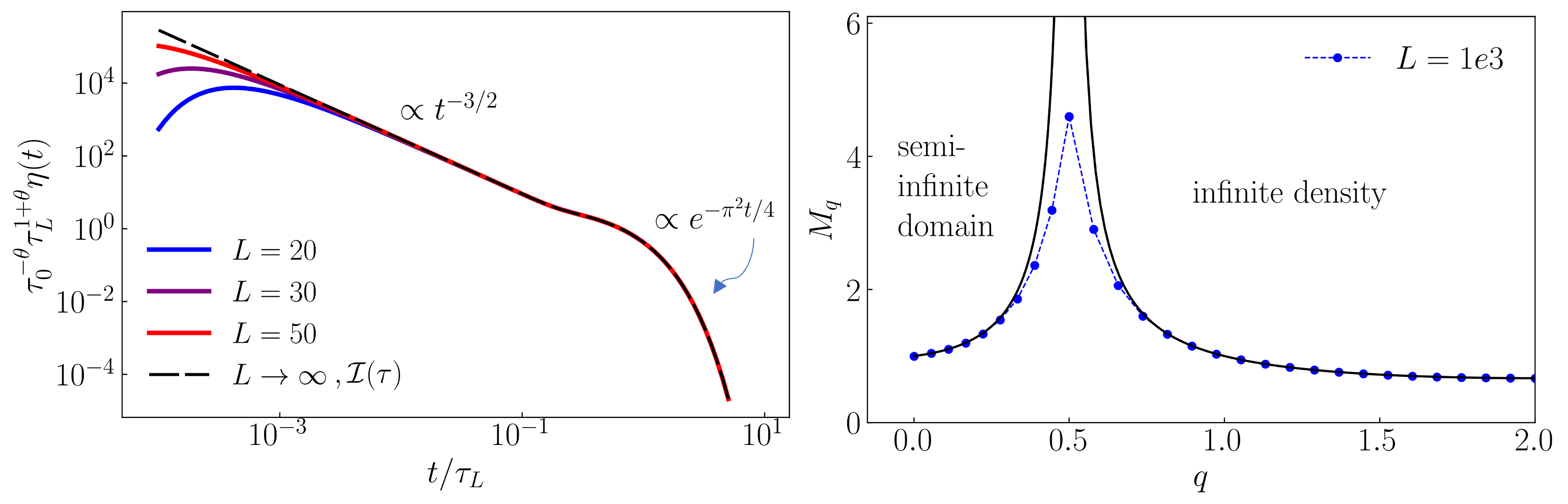}

    \caption{(a) The rescaled function $\tau_L^{3/2}\eta(t)$ for different domain sizes $L$. As the system size increases, the solutions converge towards the infinite density function $\mathcal{I}(\tau)$ given in Eq. (\ref{eq:inf1first}), represented by the black dashed line. The infinite density function captures the long-time behavior of the FPT PDF as $\tau_L$ tends to infinity. (b) Amplitude of the FPT q-moment for a particle in a one-dimensional box versus q. 
    As $\tau_L$ increases, these moments converge towards the asymptotic behavior described by Eq. (\ref{asymp1}), represented by the solid black line, and display a divergence near the critical point $q=1/2$. Note that the plot is restricted to $q\in[0,2]$, so the expected blow up of $M_q$ for $q\rightarrow \infty$ is not visible. In both panels we used $D=1/2$ and $x_0=1$.}   
    \label{fig:1dbox}
\end{figure}

 Based on the approximation of the FPT distribution, we derive the $q$th moment in the large-$L$ limit, given by (see further details in appendix \ref{appendix:1dmom})  Calculating the $q$th moment for $0 \leq q<1/2$, the integral is dominated by the $t\ll L^2/D$ region, and we have
\begin{equation}
\langle t^q \rangle \sim   \left(\frac{x_0^2}{4D}\right)^{q} \Gamma(1/2 - q)/\sqrt{\pi}.
\end{equation}
On the other hand, for $q>1/2$, the integral is dominated by the $t\ll 1$ region, so that
\begin{align}\label{momeq19}
\langle t^q \rangle &\sim \int_0^\infty dt t^q \sum_{n=0}^\infty  2Dx_0 \frac{(n+1/2)^2\pi^2}{L^3} e^{-(n+1/2)^2\pi^2 Dt /L^2} \\
&= \Gamma(q+1) \pi^{-2q} D^{-q} L^{2q}(x_0/L) \sum_{n=0}^\infty ((n+1/2)^{-2q}\\
&= \Gamma(q+1) \pi^{-2q} D^{-q} L^{2q}(x_0/L) \left(2^{2q} - 1\right) \zeta(2q)
\end{align}
Those asymptotic moments can be obtain using the scaling functions and rewrite as following
\begin{equation}\label{eq:asimp_mom}
    \langle t^q\rangle\xrightarrow{}\begin{cases}
                     \tau_0^q M_q^{-} , &  q<\frac{1}{2}\\
                     \tau_0^{\frac{1}{2}}\tau_L^{q-\frac{1}{2}} M_q^{+}, &  q>\frac{1}{2}\\
                    \end{cases} \, ,
\end{equation} 
For $q<1/2$ the moments are given by the infinite system limit, namely they do not depend on $L$, and are given by Eq. (\ref{asymp0}). Specifically 
\begin{equation}
   M_q^-=\tau_0^{-q}\int_0^\infty t^q \eta_{free}(t)dt=
                      \frac{\Gamma(1/2-q)}{4^q\sqrt{\pi}}   \, .
\end{equation}
where $\eta_{free}(t)$ is given in Eq. (\ref{eq:freesol}). On the other hand, the $q>1/2$ moments scale with the system size as $\langle t^q\rangle \propto L^{2q-1}$ and converge to the moments calculated using the infinite density function given in Eq. (\ref{eq:inf1first}), in other words,
\begin{equation}\label{asymp1}
   M_q^+=\int_0^\infty \tau^q \mathcal{I}(\tau)d\tau =
                      \pi^{-2q}2(2^{2q}-1)\zeta(2q)\Gamma(1+q) \, .
\end{equation}
We see that in the limit of large system the simple problem of diffusion in an interval exhibits two types of limiting distributions, leading to a bi-scaling behavior of the FPT moments. In this sense the infinite density is a complimentary tool to the well known Eq. (\ref{eq:freesol}). In Fig. \ref{fig:1dbox}, we observe that $M_q^{\pm}$ exhibit divergences where $q \rightarrow 1/2$ as $|q-1/2|^{-1}$. This holds as \( q \) approaches \( 1/2 \) from above and below. Additionally, in the limit \( q \to \infty \), Eq.\,(\ref{asymp1}) yields 
\(
\lim_{q\rightarrow \infty}M_q^+\rightarrow\infty \, .
\)
This can be demonstrated using the asymptotic form of the Gamma function, \( \Gamma(1+q) \sim \sqrt{2\pi q} \left( \frac{q}{e} \right)^q \) for large \( q \), and of the Riemann zeta function, \( \zeta(2q) \sim 1 \). This divergence, which arises from the Gamma function, reflecting the contribution of the long times associated with the exponential decay due to the smallest eigenvalue in Eq.\,(\ref{momeq19}), as large $q$ corresponds to rare events at the tail of the distribution.
\par It is worth to mention that similar results can be obtained for systems with two absorbing boundaries rather then one absorbing and one reflecting. In this case of the one-dimensional box, it is trivial to show that the results obtained in this section can be applied to two absorbing boundaries by taking  $L\rightarrow 2L$. 
\par  In the previous discussion, we showed that $q=1/2$ is a transition point. Now, we will demonstrate that, in general, such a critical value of $q$ exists for different models, although it may differ from the specific value of $q=1/2$. These models also demonstrate different power law tails for the FPT PDF in infinite systems. To account for these diverse diffusion scenarios, we employ scaling analysis to determine the properties of the infinite density function and its moments in a broader context. 
\section{Bi-scaling theory}\label{sec:scaling}
 To extend the case discussed in section \ref{sec:1D-diff}, we consider a random walker in a non-Euclidean space, allowing for a more general framework. In the case of anomalous diffusion, the mean square displacement (MSD) follows the relation \cite{caputo1971linear,glockle1995fractional,balakrishnan1985anomalous,barkai2001fractional}
\begin{equation}\label{eq:walkdim}
    \langle r^2 \rangle\sim \Tilde{K} t^{2/d_w},
\end{equation}
where $\Tilde{K}$ is a generalized diffusion constant and $d_w$ is the walk dimension (see table \ref{table:1}).
For a compact random walk in a finite domain, the process is characterized by two time scales. One relates to the system's size, and the other to the distance of the target from the initial location of the particle. We study the PDF of the first passage time denoted by $\eta(t)$, though first we will focus on the moments of the process. In general, we can express the scaling of the q-moment with the system's size $L$ and with the initial distance from the target $r_0$ as follows:
\begin{equation}\label{generalscal}
    \langle t^q\rangle \sim \frac{r_0^{q\mu(q)}L^{q\nu(q)}}{\tilde{K}^{d_w/2}} \, ,
\end{equation}
where $q\geq 0$. In general, $q\mu(q)$ and $q\nu(q)$ are nonlinear functions. However, assuming we have only two scales in the problem, namely that the spectrum of exponents is piecewise linear, and in the large $L$ limit:
\begin{equation}\label{generalscal1}
    q\nu(q)=\begin{cases}
                     0 , &  q<q_c\\
                    \gamma(q-q_c), &  q>q_c\\
                    \end{cases} \, .
\end{equation}
To match the dimensionality of the LHS of Eq.\,(\ref{generalscal}), for $q<q_c$ the scaling with $r_0$ must follow $q\mu(q)=d_w q$. From that, we claim in general
\begin{equation}
    q\mu(q)=d_w q-q\nu(q) \, .
\end{equation}
then in Eq.\,(\ref{generalscal1}) $\gamma =d_w$.
Below we will prove this relation in generality, however, clearly for the particle in the box problem $q_c=1/2$ and $d_w=2$. Here, $q_c$ represents the critical value of $q$ that demarcates the transition between the two scaling regions, and $d_w$ characterizes the scaling exponent $q\nu(q)$ above $q_c$. The specific values of $d_w$ and $q_c$ depend on the intrinsic properties of the system under consideration. Remarkably, we will show below that they depend on the geometry of the system. This result suggests that the early stages of the diffusion process, characterized by small values of $q$, are governed by local dynamics and do not depend strongly on the system's spatial extent. On the other hand, for $q > q_c$, the high-order moments display a power-law dependence on the system size, determined by $\gamma$. In this regime, the dynamics of the system are influenced by the finite-size effects and the overall spatial characteristics of the system become crucial.

\par We introduce a size limit to the domain by imposing a reflective wall at $r=L$, with the condition $L \gg r_0$, where $r_0$ denotes the initial distance from the target. The characteristic time scales for the first passage time problem are now defined through the scaling of the mean square displacement by $\tau_0\equiv (r_0/\sqrt{\Tilde{K}})^{d_w}$ and $\tau_L\equiv (L/\sqrt{\Tilde{K}})^{d_w}$. 
\par  For short hitting times, the first-passage PDF follows an equation of the form \cite{meyer2011universality}
\begin{equation}\label{eq:scal1}
    \eta(t)\sim \frac{1}{\tau_0}f_{\infty}(t/\tau_0)  ,
\end{equation}
where for $t\rightarrow \infty$
\begin{equation}\label{eq:scal2}
    f_{\infty}(t)\propto t^{-1-\theta} ,
\end{equation}
and $\theta$ is known as the persistence exponent. Eq. (\ref{eq:scal1}) is $L$ independent, where $f_{\infty}(.)$ corresponds to the PDF of the first passage time in an infinite system. For long times, and finite L, we anticipate a different scaling behavior. We propose a second scaling solution for the first-passage PDF, given by
\begin{equation}\label{eq:scal3}
    \eta(t)\sim \frac{1}{\tau_L}(\tau_L/\tau_0)^{\xi}\mathcal{I}(t/\tau_L),
\end{equation}
where the exact expression of $\mathcal{I}(\tau)$ is yet to be determined beyond the special example in Eq. (\ref{eq:inf1first}). Matching Eq. (\ref{eq:scal2}) and Eq. (\ref{eq:scal3}), we can obtain the exponent $\xi$. Namely for not too large t,  $\mathcal{I}(t/\tau_L) \propto (t/\tau_L)^{-1-\theta}$ in agreement with Eq. (\ref{eq:scal2}), inserting in Eq. (\ref{eq:scal3}) we find
\begin{equation}
    \eta(t)\sim \frac{1}{\tau_L}(\tau_L/\tau_0)^{\xi}(t/\tau_L)^{-1-\theta}.
\end{equation}
The finite size effect primarily influences the long-time behavior of the PDF, meaning that for short times, the PDF (as described by Eq. (\ref{eq:scal1})) remains independent of the system's size. Hence, Eq. (\ref{eq:scal2}) should be independent of $L$, and therefor independent of $\tau_L$. This indicates that $\xi$ is related to the persistence exponent and can be expressed as \begin{equation}\xi = -\theta . \end{equation}

\par To establish a scaling framework that accommodates the limit of a large system, we introduce the natural time variable $\tau\equiv t/\tau_L$. By taking the limit $L\rightarrow\infty$ and $t\rightarrow \infty$, we ensure that this rescaled time remains finite. In this limit, we define the infinite density function using the scaling solution $\mathcal{I}(\tau)$, which captures the asymptotic behavior of the first passage PDF. Specifically, using Eqs. (\ref{eq:scal1},\ref{eq:scal3}) we express it as follows:
\begin{equation}\label{eq:infdefgeneral}
\boxed{\mathcal{I}(\tau)\equiv\lim_{t,\tau_L \rightarrow \infty}\text{ }\tau_0^{-\theta}\tau_L^{1+\theta}\, \eta(t),}
\end{equation} 
For small $\tau$, to match Eq.\,(\ref{eq:scal2}), $\mathcal{I}(\tau) \propto \tau^{- 1 - \theta}$, hence the infinite density function $\mathcal{I}(\tau)$ is non-normalizable, i.e., the integral $\int_0^\infty d\tau \mathcal{I}(\tau)$ diverges. As a result, the $q$-moments, defined as $\langle t^q\rangle=\int_0^\infty t^q \eta(t)dt$, exhibit two different asymptotic behaviors, given by
\begin{equation}\label{generalscal1}
    \langle t^q\rangle\xrightarrow{}\begin{cases}
                     \tau_0^qM_q^- , &  q<\theta\\
                     \tau_L^{q-\theta} \tau_0^{\theta} M_q^+ , &  q>\theta\\
                    \end{cases} \, ,
\end{equation}
where $M_q^{\pm}$ is dimensionless and defined by 
\begin{equation}\label{eq:mqminus}
    M^{-}_q=
                     \int_0^\infty f_{\infty}(t/\tau_0)(t/\tau_0)^q d(t/\tau_0)  \, ,
\end{equation}
and
\begin{equation}\label{eq:mqplus}
    M^{+}_q=
                     \int_0^\infty \tau^q \mathcal{I}(\tau)d\tau\, .
\end{equation}
This implies a bi-scaling feature of the moments with respect to system size, as for $q<\theta$, $\langle t^q\rangle \sim \tau_L^0$, converge to the moments calculated from the first passage PDF for an infinite system $f_{\infty}(t)$ for $q<\theta$.

 On the other hand, for $q>\theta$, the moments scale with $\tau_L^{q-\theta}$ and involve the integral $\int_0^\infty \tau^q \mathcal{I}(\tau)d\tau$, which captures the contribution of the infinite density function. From Eq. (\ref{generalscal1}) we see that $q_c=\theta$, hence
\begin{equation}\label{eq:finilescale}
 \boxed{  q\nu(q)=\begin{cases}
                     0 , &  q<\theta\\
                     d_w (q-\theta), &  q>\theta\\
                    \end{cases} \, ,}
\end{equation}
and as stated above
\begin{equation}\label{generalscalmu}
   \boxed{ q\mu(q)=qd_w-q\nu(q)\, .}
\end{equation}
\par The non-analyticity of the moments amplitude at \( q_c = \theta \) can be understood by examining the moments near the transition, where they are dominated by the power-law limit of the PDF, \( \eta(t) \propto t^{-1-\theta} \). For \( q < \theta \), the dominant contribution is expected to come from the tail of $f_{\infty}(t)\propto t^{-1-\theta}$. Introducing a cutoff \( t^* \) near the divergence, from Eq.\,(\ref{eq:mqminus}) we find  
\begin{equation}
\lim_{q\rightarrow \theta} M_q^- \propto \int_{t^*}^\infty t^{q-1-\theta} dt = \frac{(t^*)^{q-\theta}}{\theta - q},
\end{equation}
while on the other hand, for \( q > \theta \), the dominant contribution comes from the short time limit of $\mathcal{I}(\tau)\propto \tau^{-1-\theta}$, giving, using Eq.\,(\ref{eq:mqplus})  
\begin{equation}
\lim_{q\rightarrow \theta} M_q^+ \propto \int_0^{\tau^*} \tau^{q-1-\theta} d\tau = \frac{(\tau^*)^{q-\theta}}{q-\theta} \,.
\end{equation}
We see that the moment amplitude generally diverges as \( M_q \propto |q - \theta|^{-1} \) when \( q \) approaches \( \theta \). This divergence is illustrated in Fig.\,(\ref{fig:1dbox}) for \( \theta = 1/2 \).

\par The infinite density function defined in Eq. (\ref{eq:infdefgeneral}), although is not a PDF due to its non-normalizability, allows us to calculate the high-order moments. For example, for a particle in a $1d$ box, this includes the mean first passage time, which in many studies is considered the most important quantifier of first passage time statistics. In this work, our main objective is to obtain explicit expressions for $\mathcal{I}(\tau)$ for various systems, employing several techniques. We next intend to validate the formulation outlined above by applying it to finite domains, particularly diffusion in a two-dimensional wedge, fractals such as the Sierpinski gasket, and a non-Markovian process (CTRW). By investigating these systems, we aim to explore the behavior of the infinite density function and assess its applicability in capturing the dynamics of first-passage processes in complex systems and fractal geometries.
\par By considering the distinct roles of the scaling functions $f_{\infty}(t)$ and $\mathcal{I}(t)$, one can obtain a more comprehensive description of the system's behavior across different time scales. Specifically, $f_{\infty}(t)$ characterizes the system's dynamics for short to intermediate times, while $\mathcal{I}(t)$ captures the behavior for intermediate to long times, as discussed earlier. Combining these two scaling functions, a uniform approximation can be derived, yielding the following expression:
\begin{equation}\label{eq:uniform22}
     \boxed{\eta(t)\sim \mathcal{N}\tau_0^{-1}f_{\infty}(t/\tau_0)\tau_L^{-1-\theta}\mathcal{I}(t/\tau_L)t^{1+\theta} \, .}
\end{equation}
Here, $\mathcal{N}$ represents a normalization factor that ensures the overall consistency of the approximation. Note that the solution depends on the walk dimension $d_w$ through the rescaled times. This smoothed approximation achieved by combining $f_{\infty}(t)$ and $\mathcal{I}(t)$ allows for a more accurate characterization of the probability density function throughout the entire time range. Further discussion of the uniform approximation will be provided in section \ref{sec:uniform}. 
\par From Eq. (\ref{eq:uniform22}), we see that the calculation of the first passage PDF for compact random walks involves the calculation of two functions, and this is a result of bi-scaling which is valid rather generally. We will show that the bi-scaling concept for first passage times holds also for the widely applicable problem of a particle diffusing in a force field, again demonstrating the generality of the approach. 
\par \textbf{Remark:} in addition to the divergence at $q=\theta$, we observe that \( M_q^{+} \) blow up as \( q \to \infty \), due to the dominance of long-time contributions associated with the smallest eigenvalue of the system. This behavior arises because higher-order moments become increasingly sensitive to rare events in the tail of the distribution. A detailed derivation using Laplace's method is provided in Appendix \ref{appendix:momamp}.

\section{Two-dimensional wedge}
Next, to verify further the generality of the above results, we consider the problem of a Brownian particle moving in a two-dimensional wedge \cite{jeon2011first,fisher1988reunions,redner2001guide,mattos2012first}, see Fig. \ref{fig:wedgedemons} for schematics. The wedge is defined between two absorbing walls at $\phi=0$ and $\phi=\pi/\nu$, with a reflective boundary at $r=L$. The diffusion Eq. in plane polar coordinates is given by
\begin{equation}\label{eq:FPwedge}
    \frac{\partial}{\partial t}P(r,\phi,t)=D\left(\partial_r^2+\frac{1}{r}\partial_r+\frac{1}{r^2}\partial_{\phi}^2\right)P(r,\phi,t) \, ,
\end{equation}
where $\phi$ is the polar angle coordinate. Given the initial condition $P(r,\varphi,t=0)=\nu \sin(\phi \nu)\delta(r-r_0)/ 2 r_0$, namely the particle starts at distance $r_0$ from the origin, the solution for an infinite system is given by \cite{redner2001guide}
\begin{eqnarray}\label{eq:wedgeinfsys}
    f_{\infty}(t) = \frac{\nu r_0}{4 t^{3/2}}e^{-r_0^2/8Dt}\left( I_{\frac{3+\nu}{2}}(\frac{r_0^2}{8 D t}) - (1-\frac{8Dt}{r_0^2}(1+\nu))I_{\frac{1+\nu}{2}}(\frac{r_0^2}{8 D t}) \right),
 \end{eqnarray}
where $I_\nu(.)$ is the modified Bessel function of the first kind. In the long-time limit, Eq.\,(\ref{eq:wedgeinfsys}) gives $f_{\infty}(t)\propto t^{-1-\nu/2}$, namely $\theta=\nu/2$. When $L$ is finite, we use the eigenfunction expansion method, and find the solution for the FPT PDF to be \cite{redner2001guide} (see appendix \ref{appendix:2dwedge})
\begin{equation}\label{eq:etawedge1}
    \eta(t)=\tau_L^{-1}\sum_n c_n k_n^2 J_{\nu}\left(k_n\sqrt{\frac{\tau_0}{\tau_L}}\right) e^{- k_n^2\frac{t}{\tau_L} } \, ,
\end{equation}
when the value of \( k \) is determined by imposing a reflecting boundary at \( r = L \), the condition of zero current, defined as \( \partial_r J_\nu(k_n r / L)|_{r=L} = 0 \). This can be expressed as:
\begin{equation}\label{eq:eigk}
    J_{\nu+1}(k_n)-J_{\nu-1}(k_n)=0  \, .
\end{equation}
 Here  
 \begin{equation}\label{eq:blabla1}
    \tau_0 = r_0^2 / D \text{ },\text{ } \tau_L = L^2 / D  \, 
\end{equation} 
are the diffusive time scales, in agreement with the more general discussion in Sec. \ref{sec:1D-diff}, namely here $d_w=2$. The coefficients $c_k$ are given by
\begin{equation}\label{eq:akwedge1}
    c_k\equiv \frac{\int_0^L r J_{\nu}(\frac{k r}{L}) dr}{\int_0^L r J_{\nu}^2(\frac{k r}{L}) dr} \, 
\end{equation}
\begin{equation}\label{eq:akwedge2}
     =\frac{2^{1-\nu} k^{1+\nu} }{(2+\nu) \Gamma(1+\nu)} \frac{ \text{}_1 F_2(1 +\frac{\nu}{2}, \{1 + \nu,2 + \frac{\nu}{2}\}, -\frac{k^2}{4})}{k J_{\nu}^2(k)+k J_{\nu-1}^2(k)- 2 \nu J_{\nu}(k)J_{\nu-1}(k)} \, ,
\end{equation}
where $_pF_q({a_1,...,a_p},{b_1,...,b_q};z) = \sum_k (a_1)_k,...,(a_p)_k / (b_1)_k,...,(b_q)_k z^k/k!$  is the generalized hypergeometric function and $(a)_k = \Gamma(a+k)/\Gamma(a)$ is Pochhammer symbol. Since both the integrals in Eq. (\ref{eq:akwedge1}) are proportional to $ L^2$, $c_k$ does not depend on $L$.
\par Under the assumptions of long time and large system, namely $t\gg\tau_0$ and $\tau_L \gg \tau_0$, the infinite density is function defined as  \begin{eqnarray}\label{ew:infden1d}
    \mathcal{I}(\tau)\equiv\lim_{t,\tau_L\rightarrow \infty}  \tau_0^{-\nu/2}\tau_L^{1+\nu/2} \eta(t) \, ,
\end{eqnarray}
so that
\begin{equation}\label{eq:infwedge1}
  \boxed{  \mathcal{I}(\tau)=\frac{2^{-\nu}}{\Gamma(1+\nu)}\sum_k c_k k^{2+\nu}  e^{-k^2 \tau} .}
\end{equation}
Here $\tau=t/\tau_L$ is the natural rescaled time. This solution is independent of the initial condition and depends on $L$ only through $\tau$. To plot this function, values of $k$ and $c_k$ are easy to obtain with a simple program \cite{wedgenote}. The short and long time limits of the infinite density function are given by
\begin{equation}\label{eq:infwedge}
    \mathcal{I}(\tau) \sim \begin{cases}
 \frac{1}{2} \Gamma(1+\frac{\nu}{2}) \tau^{-1-\nu/2}
 , &  \tau\ll 1\\
            \frac{2^{-\nu}}{\Gamma(1+\nu)} c_0 k_0^{2+\nu}  e^{-k_0^2 \tau}         , &  \tau \gg 1
                    \end{cases} \, .
\end{equation}
Notice that the short-time limit satisfies the general rule $\mathcal{I}(\tau)\propto t^{-1-\theta}$, and hence this scaling solution as a stand alone is not normalizable. The long-time behavior of $\mathcal{I}(\tau)$ is dominated by the lowest eigenvalue $k_0^2$, and it is responsible for the exponential cutoff of the PDF. $k_0$ is given by the first solution for Eq. (\ref{eq:eigk}). We now explain how to obtain the short-time limit in Eq. (\ref{eq:infwedge}). For $\tau \ll 1$, the main contribution comes from the high-order eigenvalues, meaning large $k$. In this limit, we can expand $c_k$ in Eq. (\ref{eq:akwedge1}) using the asymptotics of the hypergeometric function and the Bessel function. The result can be further simplified using trigonometric identities to obtain
\begin{equation}\label{eq:ckapproxwedge}
   c_n \simeq \frac{\pi \nu}{ k_n} \, .
\end{equation}
Plugging Eq. (\ref{eq:ckapproxwedge}) into Eq. (\ref{eq:etawedge1}), and replacing the sum by an integration over $k$ gives
\begin{equation}\label{eq:s2intwedge}
    \mathcal{I}(\tau)\rightarrow  \int_0^{\infty} dk  k^{1+\nu}  e^{- k^2 \tau } =\frac{1}{2} \Gamma(1+\frac{\nu}{2}) \tau^{-1-\nu/2} \, ,
\end{equation}  
leading to Eq.\,(\ref{eq:infwedge}). According to the principle outlined in the bi-scaling section, this short-time limit of $\mathcal{I}(\tau)$ is the same as the long time limit as found from the solution of the infinite system. At least for this example, we  can confirm this assertion, based on the solution for the infinite system given in Ref. \cite{redner2001guide}.
\par Panel (b) of Fig. \ref{fig:wedgedemons} shows the rescaled first passage time distributions for a wedge with an opening angle of $\pi/2$ and different values of the system size $L$. Clearly, the curves converge to the infinite density function $\mathcal{I}(\tau)$,though some deviations, i.e. finite size effect, are observed for small $\tau$. 

\subsection{Moments in the wedge problem}
\par
\begin{figure}[h]
    \centering
    \includegraphics[width=1\textwidth]{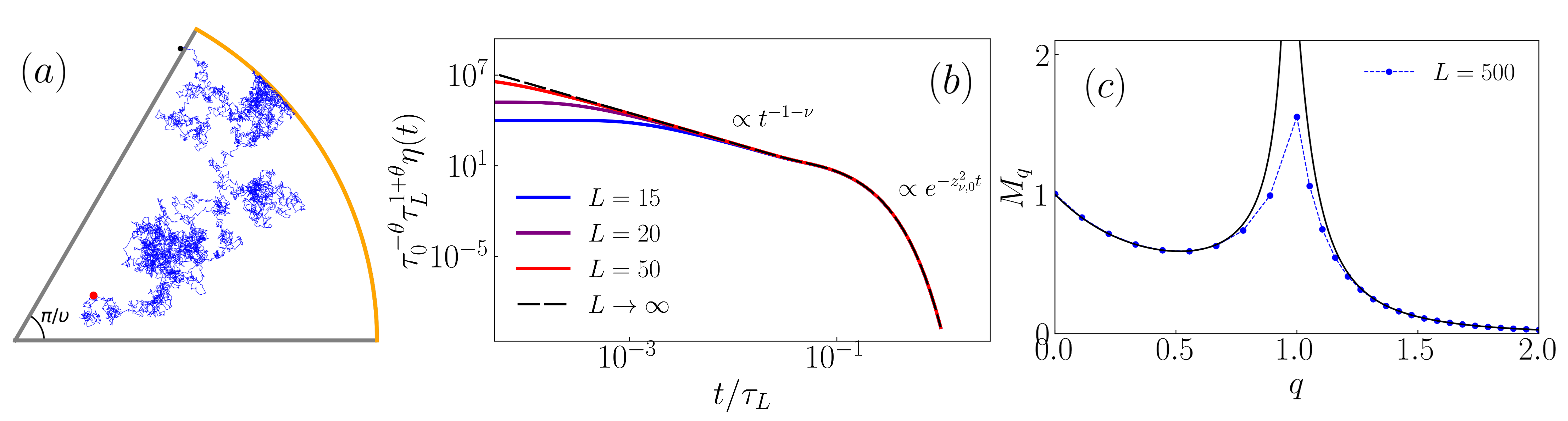}
    \caption{(a) Trajectory (in blue) of a random walker in a 2d wedge with a reflecting boundary (orange) at r=L. The lines $\phi=0$ and $\phi=\pi/\nu$ are absorbing boundaries. (b) The rescaled function $\tau_L^{1+\theta}\eta(t)$, where $\theta=\nu/2$, for different values of the domain radius $L$, where $\tau_L$ represents the characteristic time scale associated with the system size given in Eq. (\ref{eq:blabla1}) The figure demonstrates convergence towards the infinite density function given in Eq. (\ref{eq:infwedge1}), denoted by the black dashed line. The infinite density function encapsulates the long-time behavior of the FPT distribution as the domain size becomes infinitely large. (c) The moments amplitude  versus q for the FPT of a particle in a 2D wedge with an opening angle of $\pi/2$. The asymptotic moments, as found in Eqs. (\ref{m:minuswedge},\ref{m:pluswedge}) is plotted in black line. The moments for finite $L$ (blue symbols) have been calculated analytically using Eq. (\ref{eq:etawedge1}). Clearly also the finite sized system, $L=500$, exhibits a sharp transition at $q_c=1$, as theory predicts. In panels (b) and (c) we used $D=1/2$ and $r_0=1$.}
    \label{fig:wedgedemons}
\end{figure}

 The asymptotics of the moments in the large domain limit, namely $\tau_L\gg \tau_0$, are given as
\begin{equation}\label{eq:asymp_mom_wedge}
    \langle t^q\rangle\xrightarrow{}\begin{cases}
                     \tau_0^q M_q^{-}=\langle t^q\rangle_{\infty} , &  q<\frac{\nu}{2}\\
                     \tau_L^{q-\frac{\nu}{2}}\tau_0^{\frac{\nu}{2}} M_q^{+}=\tau_L^{q-\frac{\nu}{2}} \tau_0^{\nu/2}\int_0^\infty \tau^q \mathcal{I}(\tau)d\tau , &  q>\frac{\nu}{2}\\
                    \end{cases} \, .
\end{equation}
Here as mentioned the persistence exponent is $\theta=\nu/2$. For $q<\nu/2$, the moments converge to the infinite-system moments calculated using Eqs. (\ref{eq:wedgeinfsys},\ref{eq:blabla1}), given by
\begin{equation}\label{m:minuswedge}
    \langle t^q\rangle_{\infty}=\tau_0^q \frac{\nu}{2^{1+2q}}\frac{ \Gamma(1+q)\Gamma(-q+\frac{\nu}{2})}{\Gamma(1+q+\frac{\nu}{2})},
\end{equation}
these do not depend on the size of the system, $L$. In contrast, the high-order FPT moments scale with $L$ as $L^{2q-\nu}$, as Eq. (\ref{eq:asymp_mom_wedge}) indicates. Specifically, they converge to the moments calculated using the infinite density function given in Eq. (\ref{eq:infwedge}), hence
\begin{equation}\label{m:pluswedge}
   M_q^{+}= \int_0^\infty \tau^q \mathcal{I}(\tau)d\tau =  2^{-\nu}\frac{\Gamma(1+q)}{\Gamma(1+\nu)}\sum_k c_k k^{\nu-2q} \, .
\end{equation}
Notice that $M_q^{+}$ and $M_q^{-}$ are dimensionless, and as discussed in section \ref{sec:1D-diff}, these results indeed exhibit a characteristic bi-scaling behavior. The critical point indicating the transition from an infinite system to infinite density is now determined by the specific boundary conditions imposed on the system, specifically, the opening angle, namely $q_c = \nu/2$. Here, $\gamma$, as defined in Eq. (\ref{generalscal}), is $\gamma= 2$, which is the walk dimension.
In panel (c) of Fig. (\ref{fig:wedgedemons}), we plot the dimensionless parameters $M_q^{\pm}$ for $\nu=2$, meaning the critical point is $q_c=1$.
\section{CTRW}
The earlier findings presented in this work describe Markovian processes. However, diffusion in disordered systems, where a particle is hopping over barriers and between traps, is often modeled by the continuous time random walk (CTRW). The CTRW model is a non-Markovian model, and its long-time limit is described by a fractional time Fokker-Planck equation \cite{metzler1994fractional,kenkre1973generalized}, given by
 \begin{equation}\label{eq:ffp}
    \partial_{t}P(x,t)=K_{\alpha}\text{ }_{0}{D_{t}^{1-\alpha}} \partial_{x}^{2}P(x,t) ,
\end{equation}
for $0<\alpha<1$, where $K_{\alpha}$ is a generalized diffusion coefficient, and$\text{ }_{0}{D_{t}^{1-\alpha}}$ is the fractional Riemann-Liouville fractional operator, defined as
\begin{equation}
    \text{ }_{0}{D_{t}^{1-\alpha}} Z(t) = \frac{1}{\Gamma(\alpha)}\frac{\partial}{\partial t}\int_0^t dt' \frac{Z(t')}{(t-t')^{1-\alpha}} .
\end{equation}
The function $P(x,t)$ is defined in the interval $x\in[0,L]$ with an absorbing boundary condition at $x=0$ and a reflecting boundary condition at $x=L$. The solution can be obtained by the transformation \cite{barkai2001fractional} 
\begin{equation}\label{eq:p_ctrw}
    P_{\alpha}(x,t)=\int_{0}^{\infty}n(u,t)P_{1}(x,u)du ,
\end{equation}
 where $P_1(.)$ is the given solution to the regular diffusion Eq. (\ref{eq:FPeq1}), and the initial condition is $P_1(x,0) = \delta(x-x_0)$. The function $n(u,t)$ is given by
\begin{equation}\label{eq:ctrw_n}
n(u,t)=\frac{1}{\alpha}\left(\frac{K_\alpha}{D}\right)^{1/\alpha}t\, u^{-(1+1/\alpha)}\ell_{\alpha}\left(\frac{K_\alpha^{1/\alpha}t}{D^{1/\alpha}u^{1/\alpha}}\right) \, , \end{equation}
where $l_\alpha(.)$ is the one sided L\'evy probability density, defined in Laplace space as $\hat{ \ell_{\alpha}}(u) =\int_0^{\infty}\exp(-ut)l_\alpha(t)dt =  e^{-u^{\alpha}}$. The specific value of $D$, representing the diffusion coefficient for the normal diffusion Eq. (\ref{eq:FPeq1}), is irrelevant as it cancels out in the final solution for the PDF. The transformation in Eq. (\ref{eq:p_ctrw}) is related to the widely used concept of subordination; essentially it describes the effects of the fluctuations of number of jumps $u$, at time $t$, on the diffusion process in the CTRW model. The one-sided L\'evy functions $l_\alpha(.)$ is tabulated in Mathematica and its properties are well known.

\par The fractional diffusion equation describes a subdiffusive process, where the mean square displacement for a packet free of boundary conditions follows $\langle x^2 \rangle \propto t^\alpha$. Consequently, the relevant time scales are defined as 
\begin{eqnarray}\label{eq:defconstctrw}
    \tau_0=(x_0^2/K_{\alpha})^{1/\alpha}\text{ },\text{ }\tau_L=(L^2/K_{\alpha})^{1/\alpha},
\end{eqnarray}
where $x_0$ is the distance between the initial position to the absorbing point. In other words, the walk dimension is $d_w=2/\alpha$. Using Eqs. (\ref{eq:p_ctrw},\ref{eq:ctrw_n}), the FPT PDF for an infinite system is \cite{balakrishnan1985anomalous} 

\begin{equation}\label{eq:infsolctrw}
    f_{\infty}(t) = \frac{1}{\tau_0}\ell_{\alpha/2}\left(\frac{t}{\tau_0}\right) .
\end{equation}
From here it follows that, using well-known properties of stable distributions, in the limit $t\gg \tau_0$
\begin{equation} \label{eq:longtimeinfctrw}
    f_{\infty}(t)  \sim  \tau_0^{-1}\frac{\alpha}{2\Gamma(1-\alpha/2)}\left(t/\tau_0\right)^{-1-\alpha/2},
\end{equation} 
 hence the the persistence exponent is $\theta=\alpha/2$.
\par
\begin{figure}[h]
    \centering
    \includegraphics[width=1\textwidth]{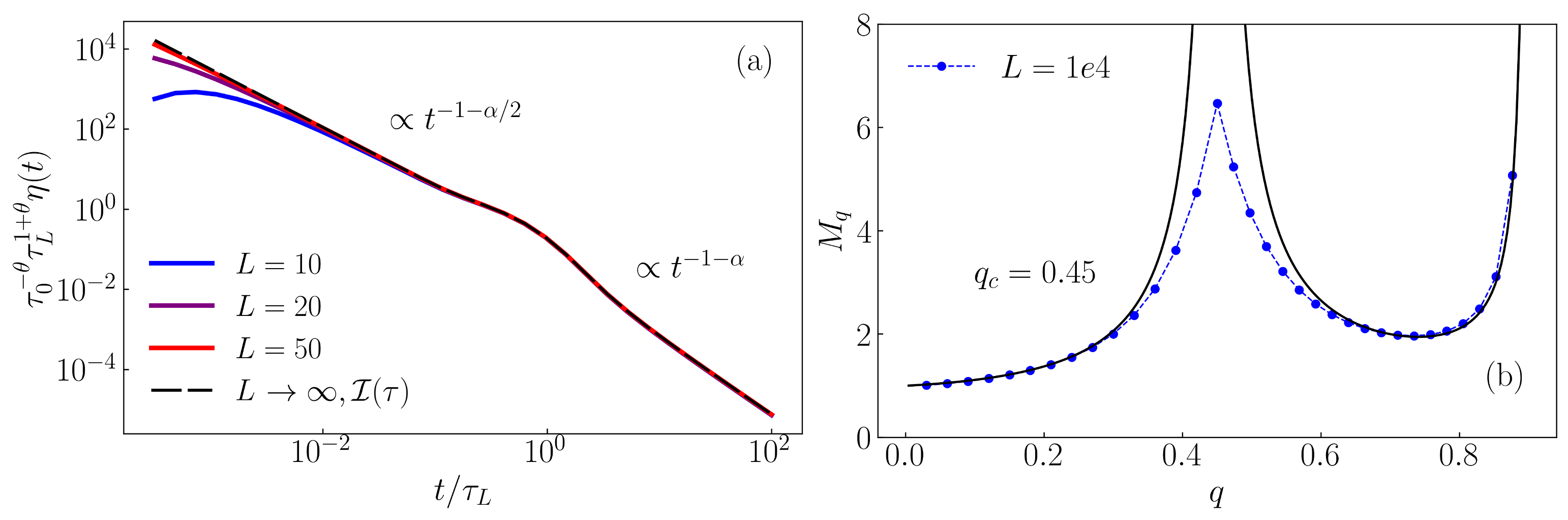}
    \caption{(a) The rescaled function $\tau_L^{1+\theta}\eta(t)$ as modeled with the fractional diffusion equation for CTRW with a waiting time exponent $\alpha = 0.9$. The plot demonstrates the convergence of the solutions as the system size, characterized by $\tau_L$, increases towards the infinite density Eq. (\ref{eq:infctrw}). (b) The asymptotic behavior of the FPT moments for CTRW with $\alpha = 0.9$. Here the critical moment is $q_c=\theta=\alpha/2=0.45$ which is clearly what we observe. Here we use $\tau_0=1$ and $K_\alpha=1$.}
    \label{fig:ctrw}
\end{figure}

 To obtain the FPT PDF for a finite domain, we applied the transformation provided in Eq. (\ref{eq:p_ctrw}) (see Appendix \ref{appendix:ctrw}). The infinite density function is then defined as
\begin{equation}\label{definfdenforctrw}
     \mathcal{I}_{\alpha}(\tau) \equiv \lim_{t,\tau_L\rightarrow \infty} \tau_0^{-\alpha/2}\tau_L^{1+\alpha/2} \eta_{\alpha}(t),
\end{equation}
where $\eta_\alpha(t)$ denoted the first-passage time PDF for CTRW, and $\tau=t/\tau_L$. Eq. (\ref{definfdenforctrw}) yields the expression:
\begin{equation}\label{eq:infctrw}
      \boxed{ \mathcal{I}_{\alpha}(\tau) =   2 \sum_n \pi^2\left(n+\frac{1}{2}\right)^2 \tau^{\alpha-1}E_{\alpha,\alpha}\left(-\pi^2\left(n+\frac{1}{2}\right)^2 \tau^{\alpha}\right),}
\end{equation}
where $E_{\alpha,\alpha}(.)$ is the Mittag-Leffler function, defined as $E_{\alpha,\alpha}(z)=\sum_{k=0}^{\infty}z^k/\Gamma(\alpha k+\alpha)$, and $\tau = t/\tau_L$. To find the $\tau \ll 1$ limit of the infinite density, we take the limit of continuous $n$. Replacing the sum with integration, Eq. (\ref{eq:infctrw}) yields (see appendix \ref{appendix:coeffctrw})
\begin{equation}\label{eq:ctrw_shortime}
\mathcal{I}_{\alpha}(\tau)|_{\tau\ll 1} \rightarrow 
 \frac{\alpha/2}{\Gamma(1-\alpha/2)}\tau^{-1-\alpha/2} ,
\end{equation}
meaning for short times, the infinite density function exhibits a power-law decay $\mathcal{I}_\alpha(\tau)\propto \tau^{-1-\alpha/2}$. Eq. (\ref{eq:ctrw_shortime}) corresponds to the long-time limit of the infinite system solution in Eq. (\ref{eq:longtimeinfctrw}). In the long-time limit ($\tau \gg 1$), applying the asymptotic form of the Mittag-Leffler function, $E_\alpha(-t^{\alpha})\sim t^{-\alpha}/\Gamma(1-\alpha)$, to Eq.\,(\ref{eq:infctrw}), we obtain
\begin{equation}\label{eq:1detatau}
   \mathcal{I}_{\alpha}(\tau)|_{\tau\gg 1}\rightarrow  \frac{2\alpha}{\Gamma(1-\alpha)} \sum_n \pi^{-2}\left(n+\frac{1}{2}\right)^{-2} \tau^{-1-\alpha} \, ,
\end{equation}
Using the sum $\sum_{n=0}^\infty (\pi(n+1/2))^{-2}=1/2$, this simplifies to 
\begin{equation}
\mathcal{I}_{\alpha}(\tau)|_{\tau\gg 1} \rightarrow 
 \frac{\alpha}{\Gamma(1-\alpha)}\tau^{-1-\alpha} ,
\end{equation}
meaning that in the long time limit, we find that $\mathcal{I}_\alpha(\tau)$ scales as $\tau^{-1-\alpha}$. Unlike the previous models discussed in this paper, the FPT PDF for the CTRW model does not exhibit an exponential cutoff for long times but rather an additional power-law decay. This behavior is illustrated in Fig. \ref{fig:ctrw}(a).

Our results demonstrate that despite the non-Markovian nature of the process (i.e. CTRW), we observe a similar bi-scaling behavior as in Markovian diffusion models. However, here the infinite density function significantly differs from its corresponding Markovian counterpart. In this case, the PDF for the FPT exhibits two distinct power-law behaviors. One arises in the short time limit, akin to the patterns observed in the previous models. Simultaneously, another power-law emerges in the long time limit, replacing the exponential tail that we observed in earlier scenarios.

\subsection{The moments}
In appendix (\ref{appendix_transformfrac}), we demonstrate that the $q$-th moment of the FPT PDF for fractional diffusion can be expressed as follows:
\begin{equation}\label{eq:ctrw_exmom1}
   \langle t^q\rangle_{\alpha} =\left(\frac{D}{K_\alpha}\right)^{q/\alpha}\langle t^{q/\alpha}\rangle _1 \frac{\Gamma[-q/\alpha]}{\alpha \Gamma[-q]}
\end{equation}
where 
\begin{equation}\label{eq:ctrw_exmom2}
\langle t^{q/\alpha}\rangle _1  =\int_0^\infty t^{q/\alpha}\eta_1(t)dt
\end{equation}
is the $q/\alpha$ moment of the FPT PDF $\eta_1(t)$, which corresponds to the solution of the ordinary diffusion equation (\ref{eq:FPeq1}). One can see from Eq. (\ref{eq:asimp_mom}) that $\langle t^{q/\alpha}\rangle _1\propto D^{-q/\alpha}$, namely the choice of $D$ is irrelevant as it will cancel out. By combining Eq. (\ref{eq:ctrw_exmom1}) with the results obtained in section \ref{sec:1D-diff}, we can express the asymptotic behavior of the moments as follows:
\begin{equation}\label{eq:asymp_ctrwmom}
    \langle t^q\rangle_{\alpha} \xrightarrow{}\begin{cases}
                     M_q^- \tau_0^{q}=\langle t^q\rangle_{\infty}, &  q<\frac{\alpha}{2}\\
                     M_q^+ \tau_L^{q-\alpha/2}=\tau_L^{q-\alpha/2}\tau_0^{\alpha/2}\int_0^\infty \tau^q \mathcal{I}_{\alpha}(\tau)d\tau , &  \frac{\alpha}{2}<q<\alpha\\
                     \infty , &  q>\alpha\\
                    \end{cases} , 
\end{equation}
where, using Eq. (\ref{eq:infsolctrw}), for $q<\alpha/2$
  \begin{equation}
   \langle t^q\rangle_{\infty}=\int_0^\infty t^qf_{\infty}(t)dt =\left(\frac{\tau_0}{4}\right)^q \frac{\Gamma(-\frac{q}{\alpha}) \Gamma(\frac{1}{2}-\frac{q}{\alpha})}{\alpha \sqrt{\pi} \Gamma(-q)}
\end{equation}
corresponds to the moments of the infinite system $f_\infty(.)$ given in Eq. (\ref{eq:infsolctrw}), and using Eq. (\ref{eq:infctrw}),
\begin{equation}
   \int_0^\infty \tau^q \mathcal{I}_{\alpha}(\tau)d\tau = 2\Gamma[1+q/\alpha] \frac{\Gamma(-\frac{q}{\alpha})}{\alpha \Gamma(-q)}\pi^{-2q/\alpha} (-1 + 2^{2q\alpha}) \zeta(2 q\alpha)
\end{equation}
are the moments calculated from the infinite density function. The time scales $\tau_0,\tau_L$ are defined in Eq. (\ref{eq:defconstctrw}). We see that the general formal Eq. (\ref{eq:finilescale}) holds with $\theta=\alpha/2$ and $d_w=2/\alpha$. We also encounter a behavior different from what we found previously, namely the FPT moments diverge for $q>\alpha$. Physically the latter divergence is expected. Based on CTRW we have power-law waiting time between jump events \cite{metzler1994fractional} and $\alpha$ order moment of the waiting time diverges. It follows that also the $\alpha$ moment of the first passage time $\langle t^q \rangle$ must diverge. In Fig. \ref{fig:ctrw}(b) using finite size system simulation, we see two diverges singular points for $q=\alpha/2$ and for $q=\alpha$.

\section{Fractal geometry}
To start out study of diffusion on fractals, we use the O’Shaughnessy-Procaccia transport equation \cite{o1985diffusion,o1985analytical,martin2011first,bray2000random}, which
describes the evolution of the propagator $P(r,t)$ of a random walker in a fractal domain
\begin{equation}\label{eq:diff_frac}
    \frac{\partial}{\partial t}P(r,t)=\frac{K}{r^{d_f-1}}\frac{\partial}{\partial r}\left( r^{1+d_f-d_w}\frac{\partial}{\partial r}\right)P(r,t). 
\end{equation}

The fractal dimension $d_f$ is defined with $M \sim r^{d_f}$ when $M$ is a volume of radius $r$ \cite{ben2000diffusion}, and $d_w$ represents the walk dimension as specified in Eq. (\ref{eq:walkdim}). We will soon check the predictions of this approach with respect to the first passage time using numerical simulations on the Sierpinski gasket. The normalization condition, determined by the fractal dimension $d_f$ is
 \begin{equation}\label{eq:FPeqsolfrac}
   \int_0^\infty P(r,t)r^{d_f-1}dr=1.
\end{equation}
\par The solution for Eq. (\ref{eq:diff_frac}) is given by the eigenfunction expansion \cite{fa2005anomalous}
 \begin{equation}\label{eq:FPeqsolfrac}
    P(r,t)=r^{(d_w-d_f)/2}\sum_n \left( a_n J_{1-\nu}(2\sqrt{\frac{\lambda_n}{d_w^2 K}}r^{d_w/2})+b_n  J_{\nu-1}(2\sqrt{\frac{\lambda_n}{d_w^2 K}}r^{d_w/2}) \right)e^{-\lambda_n t}
\end{equation}
where $\nu=d_f/d_w<1$. Imposing an absorbing boundary condition at $r=0$, meaning $P(0,t)=0$, and a non-zero current at the origin $j(r=0)\neq 0$, hence $b_n=0$. Using a reflecting boundary at $r=L$, meaning $\partial_r P(r,t)|_{r=L}=0$ therefore we find
\begin{equation}
 \lambda_n =\frac{d_w^2 }{4\tau_L}z_{-\nu,n}^2 ,
\end{equation}
Where $z_{-\nu,n}$ are the real zeros of $J_{-\nu}$. For a particle that starts at $r=r_0$ at $t=0$ \cite{fa2005anomalous}
\begin{equation}
   a_n =d_w \tau_L^{-1}\tau_0^{(1-\nu)/2}\frac{J_{1-\nu}\left(z_{-\nu,n}\sqrt{\frac{\tau_0}{\tau_L}}\right)}{J_{1-\nu}^2(z_{-\nu,n})}  ,
\end{equation}
where $\tau_L=L^{d_w}/K$ and $\tau_0=r_0^{d_w}/K$. The FPT PDF, defined as $\eta(t)=-\partial_t\int_0^L dr r^{d-1} P(r,t)$, is given by
\begin{equation}\label{eq:etafrac}
    \eta(t) = \tau_L^{(\nu-3)/2}\tau_0^{(1-\nu)/2}\frac{2^{\nu-1} d_w^2}{\Gamma(1-\nu)} \sum_{n=0}^\infty  \frac{J_{1-\nu}\left(z_{-\nu,n}\sqrt{\frac{\tau_0}{\tau_L}}\right) }{J_{1-\nu}^2(z_{-\nu,n})} 
 z_{-\nu,n}^{1-\nu} e^{-\lambda_n t}.
\end{equation}
The large-domain asymptotic solution of Eq. (\ref{eq:etafrac}) is obtained  by Meyer et al \cite{meyer2011universality} to be 
\begin{equation}\label{eq:fptfractal}
    \eta(t) \sim \frac{2^{2\nu-3} \Gamma(\nu) }{\Gamma(2-\nu)}\tau_L^{\nu-2}\tau_0^{1-\nu} d_w \sum_{n=0}^{\infty}  \frac{J_{\nu}(z_{-\nu,n}) }{J_{1-\nu}(z_{-\nu,n})}  z_{-\nu,n}^{3-2\nu}  e^{-d_w^2 z_{-\nu,n}^2 t/4\tau_L}.
\end{equation}
This can be obtained from Eq. (\ref{eq:etafrac}) using the asymptotic of the Bessel function $J_\alpha(z)\sim (z/2)^\alpha / \Gamma(1+\alpha)$ for $z\ll \sqrt{1+\alpha}$, and the identity $J_{\nu-1}(z)J_{-\nu}(z)+J_{1-\nu}(z)J_{\nu}(z)=\frac{2}{z \Gamma(1-\nu) \Gamma(\nu)}$, where $J_{-\nu}(z_{-\nu})=0$.
The infinite density is then defined as
\begin{equation}\label{eq:defidfrac}
     \mathcal{I}(\tau) \equiv \lim_{t,\tau_L\rightarrow \infty} \tau_0^{-1+\nu}\tau_L^{2-\nu} \eta(t).
\end{equation}
then from Eqs. (\ref{eq:fptfractal},\ref{eq:defidfrac})
\begin{equation}\label{eq:fptfractalinf}
   \boxed{ \mathcal{I}(\tau)= \frac{2^{2\nu-3} \Gamma(\nu) }{\Gamma(2-\nu)} d_w \sum_{n=0}^{\infty}  \frac{J_{\nu}(z_{-\nu,n}) }{J_{1-\nu}(z_{-\nu,n})}  z_{-\nu,n}^{3-2\nu}  e^{-d_w^2 z_{-\nu,n}^2 \tau/4}.}
\end{equation}
A pointed out by Meyer et al. \cite{meyer2011universality}, this solution is not integrable due to its small $\tau$ limit, as in other examples in this paper. 
\par
\begin{figure}[h]
    \centering
    \includegraphics[width=1\textwidth]{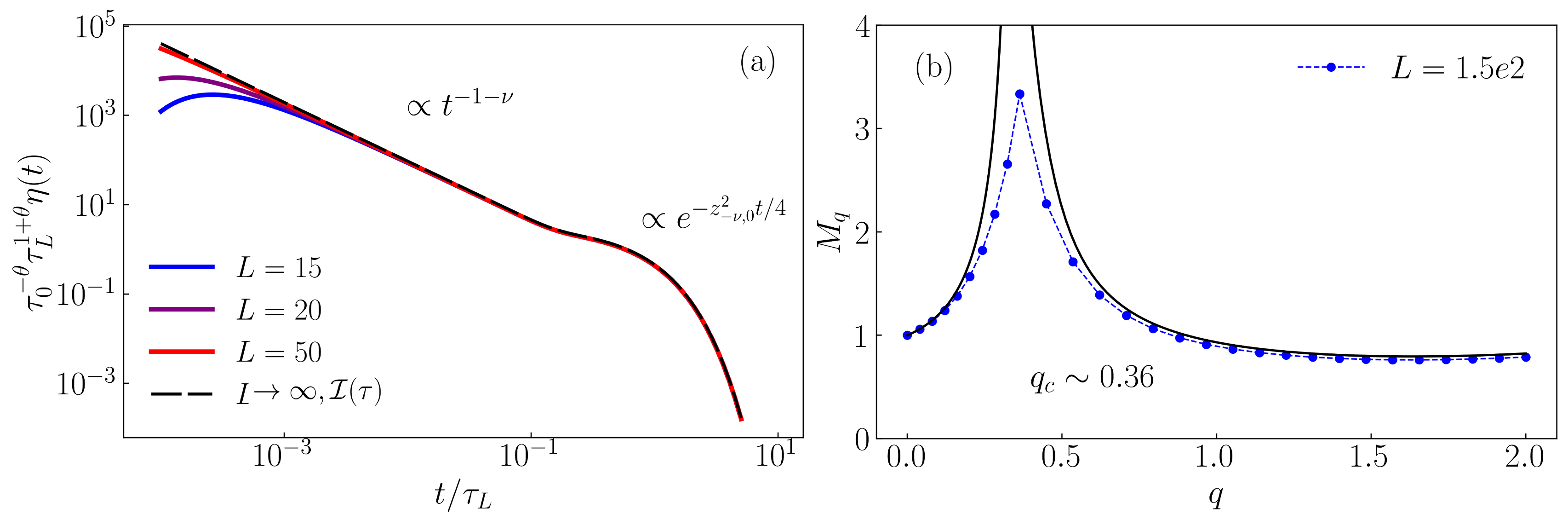}
    \caption{ (a) The rescaled function $\tau_L^{1+\theta}\eta(t)$ for O’Shaughnessy-Procaccia diffusion Eq. (\ref{eq:diff_frac}) with $\nu = 0.64$. (b) The asymptotic behavior of the FPT moments with $\nu = 0.63$.  Here we took $\theta=q_c\sim 0.36$, $K=1$ and $\tau_0=1$.}
    \label{fig:fracfig}
\end{figure}

 Let us study the exact form of divergence of the infinite density in Eq. (\ref{eq:fptfractalinf}) for small $\tau$. To study the short time limit, we replace the summation with integration, and obtain
\begin{equation}
    I(\tau)\rightarrow \frac{ d_w^2 }{ \Gamma(\nu) \Gamma(2-\nu)}  \int_0^{\infty} dk   k^{3-2\nu}  e^{-d_w^2 k^2 \tau/4}
\end{equation}
\begin{equation}
    = \frac{d_w^{-2+2\nu} }{2 \Gamma(\nu) } \tau^{-2+\nu} ,
\end{equation}
which is non-integrable, since $0<\nu < 1$. This corresponds to the long-time limit of the solution for the FPT in an infinite system, which was studied by Kwok at al \cite{fa2005anomalous}
\begin{equation}\label{eq:infsolfractal}
    f_{\infty}(t) =\frac{d_w^{-2+2\nu} }{2 \Gamma(\nu) } e^{-\frac{\tau_0}{d_w^2t}}( t/\tau_0)^{-2+\nu}  /\tau_0.
\end{equation}
Here clearly the persistence exponent is $\theta=1-\nu$.
Using the solution provided in Eq. (\ref{eq:fptfractal}) and the exact solution (see Appendix \ref{appendix:fractal}), the asymptotic behavior of the moments in the large domain limit in given by
\begin{equation}\label{eq:fractalmom}
    \langle t^q\rangle\xrightarrow{}\begin{cases}
                    M_q^-= \langle t^q\rangle_{\infty} , &  q<1-\nu \\
                    \tau_L^{q-(1-\nu)} \tau_0^{1-\nu}M_q^+= \tau_L^{q-(1-\nu)} \tau_0^{1-\nu} \int_0^\infty \tau^q \mathcal{I}(\tau)d\tau , &  q>1-\nu \\
                    \end{cases} \, .
\end{equation}
Where $\langle t^q\rangle_{\infty} $ is obtained using Eq. (\ref{eq:infsolfractal})
\begin{eqnarray}
    \langle t^q\rangle_{\infty} =\tau_0^qd_w^{-2q}\frac{ \Gamma(1-\nu-q)}{2 \Gamma(\nu) }  .
\end{eqnarray}
As illustrated in Fig. \ref{fig:fracfig}(b), we observe distinct behaviors depending on the value of $q$ relative to the critical threshold $1-\nu$. When $q$ falls below this threshold, the system behaves similarly to an infinite system. However, when $q$ exceed the threshold, the moments exhibit scaling behavior with the system size, characterized by the term $\langle t^q\rangle \propto L^{d_w(q-(1-\nu))}$ in Eq. (\ref{eq:fractalmom}). 
From Eq. (\ref{eq:fractalmom}), the validity of the formula for bi-linear scaling, Eq. (\ref{eq:finilescale}) is easy to verify.
\subsection{The Sierpinski gasket}
Previous work showed that the O’Shaughnessy-Procaccia equation describes some statistical properties of random walks and first passage times on different fractal geometries \cite{meyer2011universality,zunke2022first}. 
Here, a simulation was run to explore the characteristics of a discrete time random walk on the Sierpinski gasket, a deterministic fractal extensively used as a toy model in the field of statistical physics (see appendix \ref{appendix:simulation}). This model, known for its intricate self-similar structure, serves as a compelling model for investigating various phenomena and properties within the realm of random walks \cite{fa2003power,benichou2008zero,hambly1998stochastic,hambly1992brownian,hambly1997brownian}.
\par
\begin{figure}[h]
    \centering
    \includegraphics[width=0.8\textwidth]{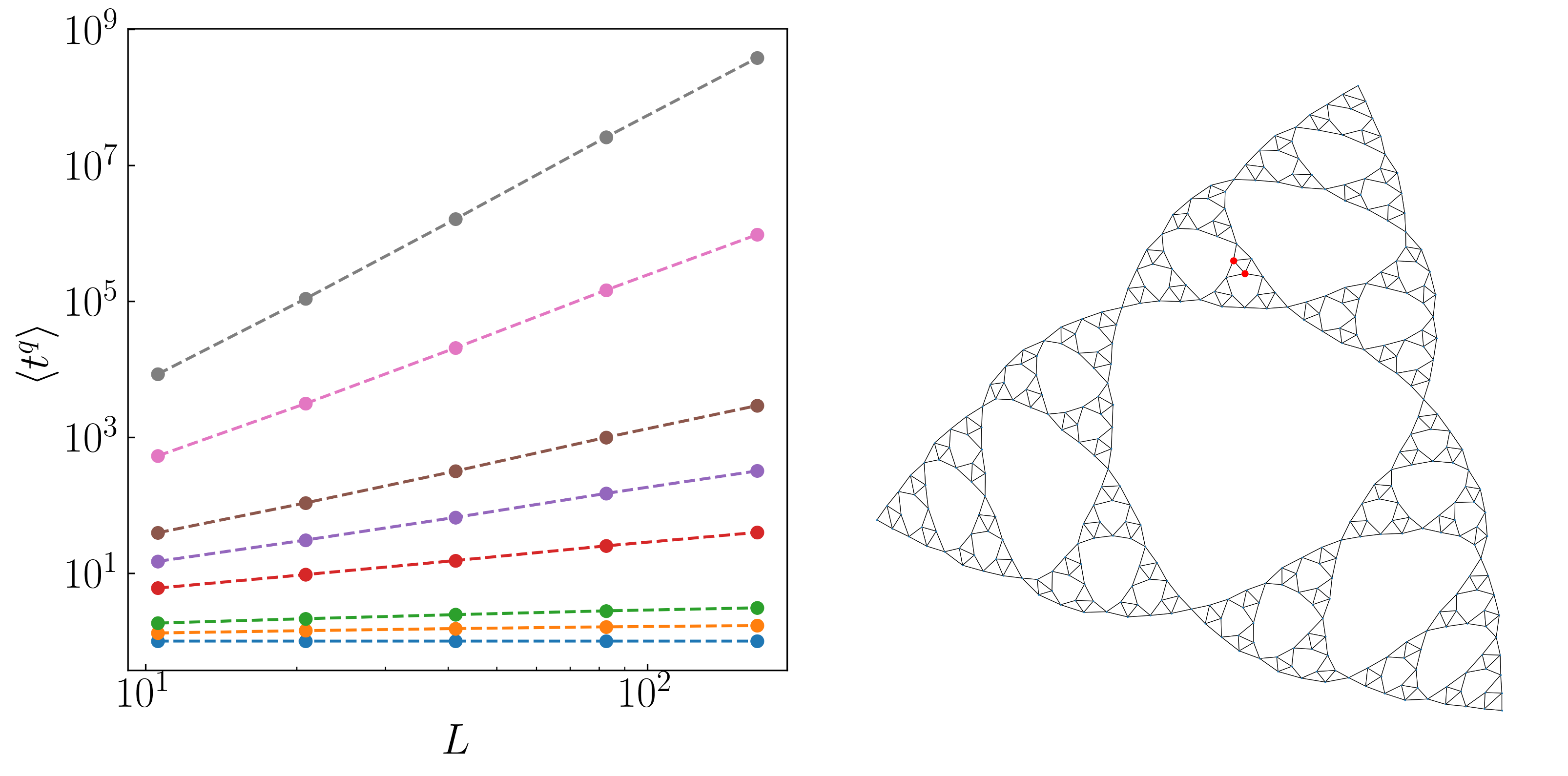}
    \caption{Left: The moments $\langle t^q \rangle$ versus $L$, for $q=0,0.2,0.3,0.6,0.8,1,1.5,2$, from bottom to top. From here we can estimate the spectrum of the exponent defined with $\langle t^q \rangle\sim L^{q\nu(q)}$ as defined in Eq. (\ref{eq:finilescale}). The low order moments, $q<q_c$ show saturation effect, while for $q> q_c$ the moments increase with system size, as described by the scaling relationship in Eq. (\ref{eq:fractalmom}). The system's size $L$ is given by $L=N^{1/d_f}$, where N is the number of nodes in the graph, hence the scaling parameter $\tau_L$ is given by $\tau_L=N^{d_w/d_f}$. Right: Order-5 Sierpinski gasket, the red dots represent the initial condition and the target.}
    \label{fig:sirp}
\end{figure}

\begin{figure}[h]
    \centering
    \includegraphics[width=1\textwidth]{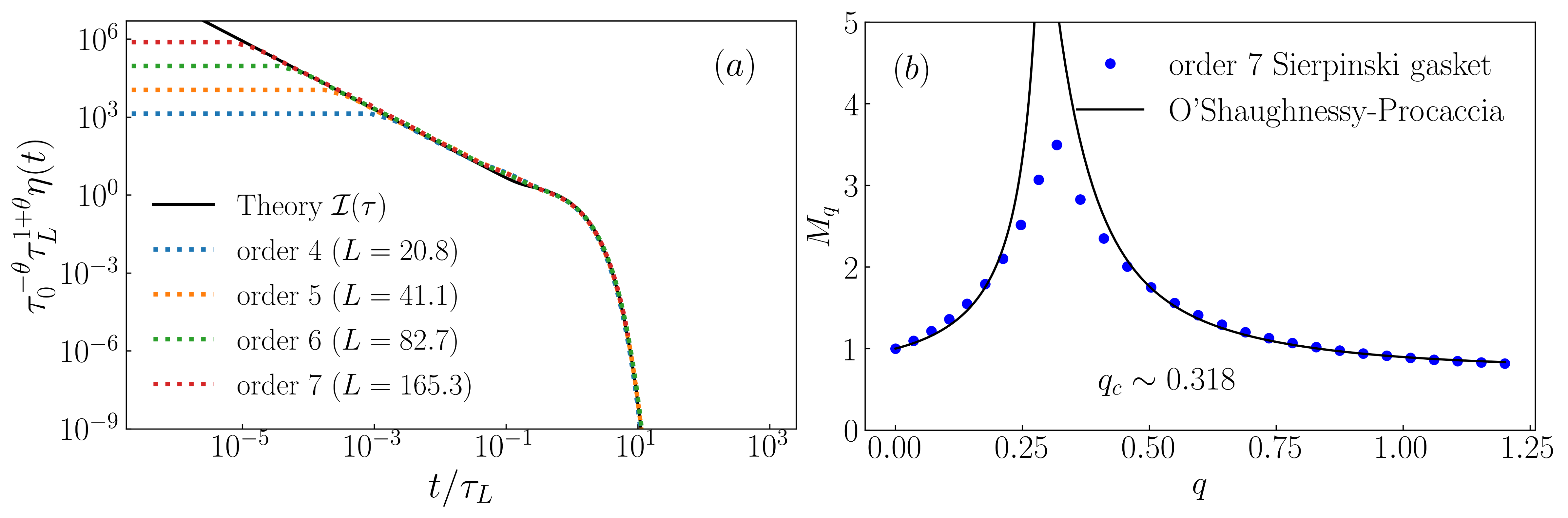}
    \caption{(a) Simulation of the rescaled FPT PDF on a Sierpinski gasket for different fractal orders approched in the limit of large order TO the infinite density for the O’Shaughnessy-Procaccia model given in Eq. (\ref{eq:fptfractalinf}) (solid black line).
Notice the short-time power law behavior $\sim t^{-2+\nu}$, for as the order of the system grows, the function converges toward the expected limit. 
    (b) The moments of the FPT on an order-7 Sierpinski gasket compared with the asymptotic moments for the O’Shaughnessy-Procaccia model given in Eq. (\ref{eq:fractalmom}). We see that $M_q$ exhibit a clear transition for $q_c=1-\nu\sim 1.318$, similar to the other examples in this paper. Finite size  deviations are also clearly observed in the vicinity of $q_c$. }
    \label{fig:sirp2}
\end{figure}

 The simulation results have been rescaled to be
compared to the theoretical expression given in Eq. (\ref{eq:fptfractal}). We chose $K=1$, while the fractal dimension and the walk dimension of the Sierpinski gasket are given by
\begin{equation}
    d_f=\log(3)/\log(2)\text{ },\text{ }d_w=\log(5)/\log(2),
\end{equation}
correspondingly, hence $\theta=1-\log(3)/\log(5)$. On a fractal we use $L = N^{1/d_f}$ where $N$ is the mass, namely the number of nodes in the simulation. The relation between the number of nodes $N$ and the fractal order $m$ (that is, the number of iterations or recursive steps taken to generate the fractal pattern) for the Sierpinski gasket is given in general by $N=(3 + 3^{m+1})/2$ \cite{teguia2005sierpi}. To ensure consistency with the large-system limit, the source and target points were deliberately selected to be in close proximity to each other (see Fig. \ref{fig:sirp}). In Figure \ref{fig:sirp2}(a), as the fractal order in the simulation (and thus the system's size) increases, we observe a convergence towards an infinite density function for the O’Shaughnessy-Procaccia model given in Eq. (\ref{eq:fptfractalinf}). Additionally, the simulation demonstrates our predicted bi-scaling of the moments. As seen in Fig. \ref{fig:sirp}, while the low-order moments show no apparent dependence on the system's size, the high-order moments are clearly increasing with the system size. These findings strongly suggest the presence of two scaling solutions, consistent with the expectations outlined in our study. Fig. \ref{fig:sirp2} demonstrates an excellent agreement between the numerical results and the theoretical prediction given in Eq. (\ref{eq:fptfractal}), supporting the validity of the theoretical model in describing the system's behavior.  

\section{FPT for particles in a force field}
Next, we consider the case of an overdamped particle in one dimension subjected to an attracting force field $F(x)$, with the initial condition $P(x,0)=\delta(x-x_0)$ and an absorbing boundary at the origin $x=0$, so that $P(0,t)=0$. The main condition is that the force is binding, hence the particle will reach the origin with probability one, the search of the target is a compact search. When the temperature is high, we expect that the effect of the force diminishes, causing the FPT probability density to resemble that of a free particle with some modifications. In the short-time limit, we anticipate observing the free-particle solution described in Eq. (\ref{eq:freesol}), featuring a fat-tailed power law decay followed by a cutoff in the long-time limit. Unlike the previous examples, a reflecting boundary at a finite position is no longer necessary, as the potential prevents the particle from escaping to infinity. In the following cases, the bi-scaling behavior of the PDF relates to the transition from force-free particle behavior to the infinite density solution associated with the external force. Here we will define the length scale $L$ differently than the previous models, see below. The Fokker-Planck equation describe the particle's movement is given by
\begin{equation}\label{eq:FPeq2}
\frac{\partial}{\partial t}P(x,t) = \tilde{\mathcal{L}}_{FP} P(x,t),
\end{equation}
where the Fokker-Planck operator \(\tilde{\mathcal{L}}_{FP}\) is defined as
\begin{equation}\label{eq:FPop}
\tilde{\mathcal{L}}_{FP} = D\left\{\frac{\partial^2 }{\partial x^2} + \frac{1}{T}\frac{\partial}{\partial x}\left(\frac{\partial V(x)}{\partial x}\right)\right\},
\end{equation}
and $V(x)$ is the external potential generating the fore, $T$ is the temperature of the reservoir, $D$ is the diffusion constant, and the Boltzmann constant is set to one. We consider here purely attractive potentials so that deterministically, the particle will always be drawn toward the absorbing boundary at zero. The solution for Eq. (\ref{eq:FPeq2}) is defined in the semi-infinite domain, $x\in[0,\infty)$. The particle starts at $x=x_0$ at time $t=0$.

\subsection{Constant force}
For a linear potential, meaning $F(x) =  -F $, with $F\geq 0$, the FPT PDF is given by \cite{schrodinger1915theorie}
\begin{equation} 
    \eta(t)=\frac{x_0}{2\sqrt{D\pi}}t^{-3/2}e^{\frac{-(x_0-D\frac{F}{k_B T}t)^2}{4Dt}} \, .
\end{equation}
In this problem, we have two competing time scales. First, $\tau_0=x_0^2/D$ is the typical time for a force-free particle to reach the origin. Second is the time scale associated with the external force, given by $\tau_L=L^2/D$, where $L$ is the confining length defined as $L\equiv T/F$, so that $\tau_L=T^2/(F^2D)$. Let us rewrite the FPT PDF in terms of both time scales as

 \begin{equation}\label{eq:linsol1}
    \eta(t)=\frac{1}{\tau_0\sqrt{4\pi}}(t/\tau_0)^{-3/2}\exp\left(-\frac{(\sqrt{\tau_0/\tau_L}-t/\tau_L )^2 }{4(t/\tau_L)}\right) \, .
\end{equation}
The moments of the FPT, defined as $ \langle t^q\rangle=\int_0^{\infty} t^q \eta(t) dt$, calculated from Eq. (\ref{eq:linsol1}) are
 \begin{equation}\label{linmomex}
    \langle t^q\rangle=\frac{\tau_0^{\frac{q}{2}+\frac{1}{4}}}{\sqrt{\pi}} \tau_L^{\frac{q}{2}-\frac{1}{4}} e^{\sqrt{\frac{\tau_0}{4\tau_L}}} K_{\frac{1}{2}-q}\left(\sqrt{\frac{\tau_0}{4\tau_L}}\right)\, ,
\end{equation}
where $K_{1/2-q}(.)$ is the modified Bessel function of the second kind. From Eq. (\ref{eq:linsol1}), we see that $\tau_0$ controls the short-time behavior, where the PDF behaves as the free-diffusion solution (\ref{eq:freesol}). We focus our interest on the limit of a weak force, or high temperature, that is $\tau_0\ll \tau_L\rightarrow x_0 F/ T \ll 1$, meaning the typical work it takes to move the particle from the initial point to the origin, is much smaller than the thermal energy. In this limit, and for long times $t\gg \tau_0$, Eq. (\ref{eq:linsol1}) reduces to $\eta(t)\propto t^{-3/2}\exp(-t /4 \tau_L)$. Expanding $\langle t^q\rangle$ in the weak force limit, the Bessel function asymptotics is given by $K_n(z)\sim z^n(2^{-1-n}\Gamma(-n)+O(z^2))+z^{-n}(2^{-1+n}\Gamma(n)+O(z^2))$. We have two cases, either the first term dominates or the second, depending on whether $q>\theta$ or $q<\theta$. Here  $\theta=1/2$, therefore the solution separates into two different cases as 
\begin{equation}\label{eq:linmom}
    \langle t^q\rangle \simeq \begin{cases}
                     \frac{4^{-q}}{\sqrt{\pi}}\Gamma(\frac{1}{2}-q)\tau_0^q , &  q<\frac{1}{2}\\
                     \frac{4^{q-1}}{\sqrt{\pi}}\Gamma(q-\frac{1}{2})\sqrt{\tau_0} \tau_L^{q-\frac{1}{2}} , &  q>\frac{1}{2}\\
                    \end{cases}.
\end{equation}
As seen before, the $q<1/2$ moments are identical to the moments of the force-free diffusion Eq.\,(\ref{asymp0}), which indeed only exist for $q<\frac{1}{2}$. The high-order moments ($q>1/2$) in Eq. (\ref{eq:linmom}) are sensitive to the strength of the force. Using the FPT denoted $t$, we define the random variable $\tau=t/\tau_L$, so that we can write the $q>1/2$ moments as $\tau_L^{q-1/2}\langle \tau^q\rangle$, where $\langle \tau^q\rangle=\int d\tau \tau^q\mathcal{I(\tau)} $. To find the infinite density function $\mathcal{I}(\tau)$, we can use directly Eq.\,(\ref{eq:linsol1}), consider the limit
\begin{eqnarray}\label{eq:infdendefforce}
    \mathcal{I}(\tau)\equiv \lim_{\substack{t,\tau_L\rightarrow \infty \\ t/\tau_L \rightarrow \text{fixed}}}  \tau_0^{-1/2}\tau_L^{3/2} \eta(t) \, .
\end{eqnarray}
which gives
\begin{equation}\label{eq:infdenlin}
\boxed{\mathcal{I}(\tau) = \sqrt{\frac{1}{4\pi}} \tau^{-3/2}e^{-\frac{\tau}{4}} \, ,}
\end{equation}
which clearly provides the $q>1/2$ moments as given in Eq.\,(\ref{eq:linmom}). We showed in the first section that for a finite system, the cutoff resulting from the reflective boundary ensures the existence of all moments, and, as we saw here, so does an external bias. Similarly to section 2, the exponent $\theta=1/2$ determines the critical value of $q$, separating the free-diffusion region and the infinite density region, so the larger $\tau_L$ we take, the closer the exact moments are to that asymptotic given by
\begin{equation}\label{eq:momlin}
    \langle t^q\rangle\xrightarrow{}\begin{cases}
                     \tau_0^q M_q^{-}=\langle t^q\rangle_{\mathrm{free}} , &  q<\frac{1}{2}\\
                     \tau_0^{\frac{1}{2}}\tau_L^{q-\frac{1}{2}}M_q^+=\tau_0^{\frac{1}{2}}\tau_L^{q-\frac{1}{2}} \int_0^\infty \tau^q \mathcal{I}(\tau)d\tau , &  q>\frac{1}{2}\\
                    \end{cases},
\end{equation}
with $M_q^- = 4^{-q}\Gamma(1/2-q)/\sqrt{\pi}$ and $M_q^+=4^{q-1}\Gamma(q-\frac{1}{2})/\sqrt{\pi}$. Here the high-order moments ($q>1/2$) correspond to the moments calculated using the infinite density given in Eq. (\ref{eq:infdenlin}), and $\langle t^q\rangle_{\mathrm{free}}$ are the moments that correspond to the force-free diffusion, given in Eq. (\ref{asymp0}). We see from Eq. (\ref{eq:linmom}) that both terms diverge where the Gamma function argument approaches zero at the critical moment $q\rightarrow 1/2$.

\subsection{Harmonic potential}
For a particle subject to a harmonic potential we denote the potential as $V(x)= \kappa x^2/2$, and the FPT PDF is given by (see appendix \ref{appendix:force})
\begin{equation}\label{eq:etahar1}
    \eta(t)= \sum_{n=0}^\infty \frac{\tau_L^{-1}}{\Gamma(\frac{1}{2}-n)\Gamma(2n+1)}  H_{2n+1}(\sqrt{\frac{\tau_0}{2\tau_L}})e^{-(2n+1) t/\tau_L}\, ,
\end{equation}
where $H_{2n+1}(\cdot)$ are the Hermite polynomials, and \(\tau_0 = x_0^2/D\) is the diffusive time scale as defined in section \ref{sec:1D-diff}. This result follows from the general solution outlined in \cite{risken1996fokker}. In our case, due to the absorbing boundary condition at the origin, only the odd Hermite polynomials are used in the expansion. The confinement length is defined as \(L=\sqrt{T/\kappa}\), meaning the characteristic time scale associated with the force is \(\tau_L = L^2/D = T/(D \kappa)\). This corresponds to the time scale that controls the long-time behavior of the FPT. In the high temperature limit $x_0/L\ll 1$, we expand $H_{2n+1}(z)\sim \sqrt{\pi}2^{2n+1}(2n+1)/\Gamma(1/2-n)$. In that limit, the PDF $\eta(t)$ in Eq.\,(\ref{eq:etahar1}) gives
\begin{equation}\label{eq:etaharaprox}
    \eta(t)\simeq \sqrt{\frac{\pi \tau_0}{2\tau_L^3}} \sum_{n=0}^\infty \frac{(2n+1)2^{2n+1}}{\Gamma(\frac{1}{2}-n)^2\Gamma(2n+1)} e^{-(2n+1)t/\tau_L}\, .
\end{equation}
Using the re-scaled time $\tau\equiv t/\tau_L$, the infinite density is defined to be $\mathcal{I}(\tau)\equiv \lim_{t,\tau_L\rightarrow \infty}\tau_0^{-1/2}\tau_L^{3/2}\eta(t)$. Performing the summation over n, we obtain
\begin{equation}\label{eq:infho}
\boxed{\mathcal{I}(\tau)= \sqrt{\frac{2}{\pi}}\frac{4 e^{-\tau}}{(1-e^{-2\tau})^{3/2}}\, .}
\end{equation}
The asymptotics of Eq.(\ref{eq:infho}) is given by 
\begin{equation}
\mathcal{I}(\tau) \sim \begin{cases}
 \tau^{-3/2}  /\sqrt{\pi}
 , &  \tau\ll 1\\
            4 e^{- \tau}  /\sqrt{2\pi}        , &  \tau \gg 1
                    \end{cases}\, .
\end{equation}
For large times, the FPP is often approximated by taking the leading term in the sum from Eq. (\ref{eq:etaharaprox}), which gives an exponential cutoff. That approximation, although valid for the long time limit, provides a very partial part of the distribution, and a partial description of large $\tau$. For short times, Eq. (\ref{eq:infho}) converge to the power-law decay as the force-free solution.
\par The moments of the FPT, defined as $ \langle t^q\rangle=\int_0^\infty t^q \eta(t) dt$, when $q>0$, are calculated from Eq. (\ref{eq:etahar1}) as
\begin{equation}\label{eq:harmomex11}
     \langle t^q\rangle=\Gamma(1+q)\tau_L^q\sum_{n=0}^\infty \frac{(2n+1)^{-1-q}}{\Gamma(1/2-n)\Gamma(2n+1)}  H_{2n+1}\left(\sqrt{\frac{\tau_0}{2\tau_L}}\right) \, .
\end{equation}
Expanding Eq. (\ref{eq:harmomex11}) in the limit of a weak force, or $\tau_L \gg \tau_0$, the moments converge to the asymptotic result as given in Eq. (\ref{eq:momlin}), where the low-order moments are given by the force-free moments, given in Eq. (\ref{asymp0}), and the high-order moments ($q>1/2$) are calculated from the infinite density function, for example
\begin{equation}\label{eq:spesificmomho}
    \int_0^\infty \tau^q \mathcal{I}(\tau)d\tau={}\begin{cases}
    \frac{1}{2\sqrt{\pi}}\Gamma(q-1/2) , &  q\rightarrow 1/2\\
                     \sqrt{\pi/2} , &  q=1\\
                    \sqrt{2\pi}\log{2}, &  q=2 \\
                    \end{cases},
\end{equation}
hence as $q\rightarrow 1/2^+$ the q-moment of the FPT diverges. For a general value of $q$ this integral is not analytically solvable, nevertheless, we found an approximated expression given by 
\begin{equation}\label{eq:approxmomho}
    \int_0^\infty \tau^q \mathcal{I}(\tau)d\tau \simeq \frac{\Gamma(1+q)}{\sqrt{2\pi}}\left(2+\frac{2^{1-q}}{\sqrt{\pi}}\zeta\left(\frac{1}{2} + q, \frac{3}{2}\right) \right) 
\end{equation}
where $\zeta(\cdot)$ is the zeta function. This approximation is obtained by using the infinite-sum form of the infinite density in Eq.\,(\ref{eq:etaharaprox}) and take the large-$n$ limit, beside of the $n=0$ term. In the limit $q\rightarrow 1/2$ Eq.\,(\ref{eq:approxmomho}) gives $\int_0^\infty \tau^q \mathcal{I}(\tau)d\tau\rightarrow (q-1/2)^{-1}/(2\sqrt{\pi})$, which is consistent with the asymptotics of Eq.\,(\ref{eq:spesificmomho}).
\par
\begin{figure}[h]
    \centering
    \includegraphics[width=1\textwidth]{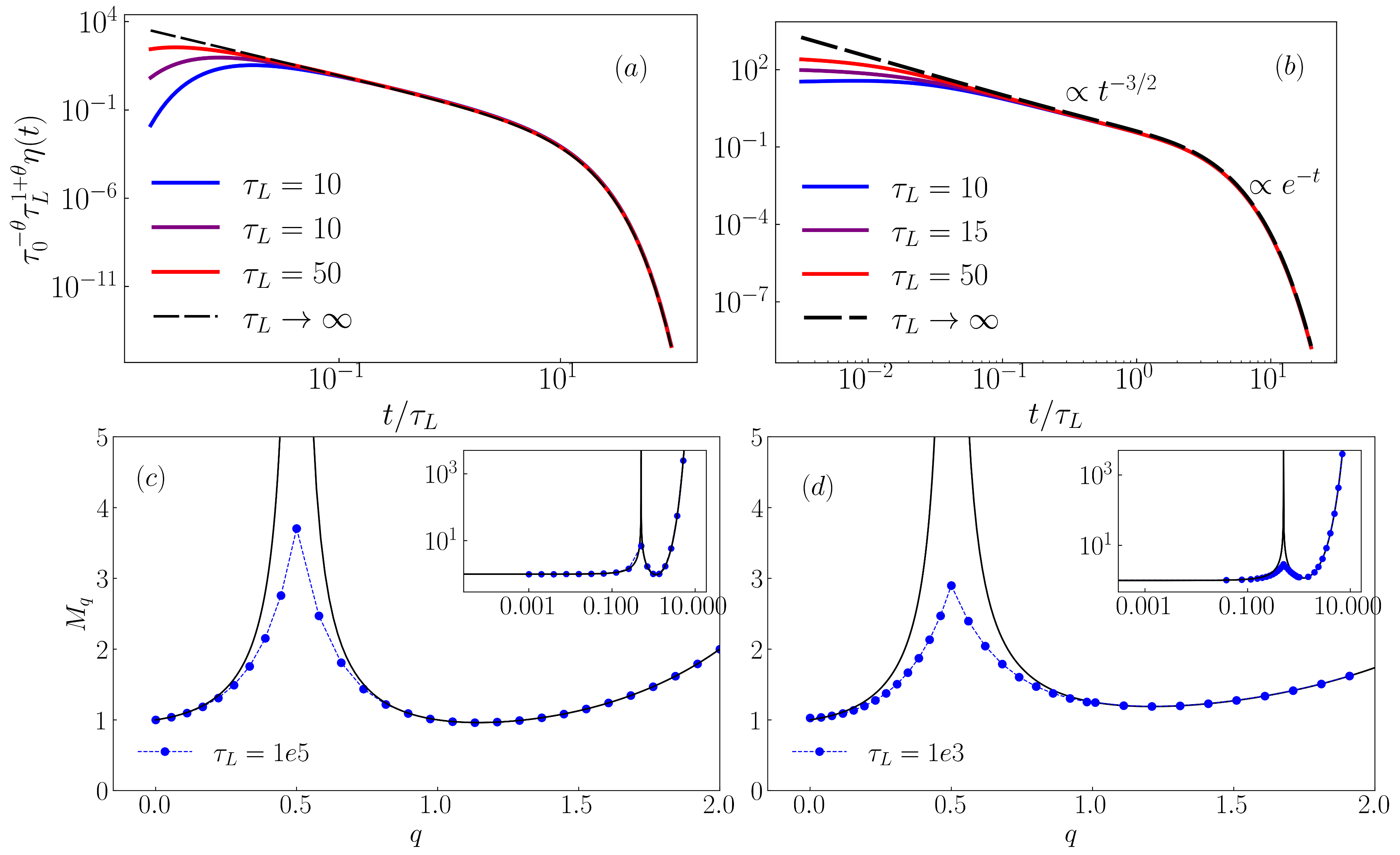}
    \caption{ 
 The rescaled FPT PDF $\tau_L^{3/2}\eta(t)$, for different values of $\tau_L$ for the linear potential Eq.\,(\ref{eq:infdenlin}) (a) and the Harmonic potential Eq.\,(\ref{eq:infho}) (b). In this context, $\tau_L$ serves as a measure of the potential strength rather than representing the system size. The moments $M_q$ in Eq.\,(\ref{eq:momlin}) for the FPT of a particle in a linear potential (c) and a Harmonic potential (d). The moments converge to their asymptotic values while $\tau_L$ increases, meaning the strength of the potential decreases. In the insets of panels (c) and (d), the behavior of $M_q$ is shown up to $q=10$ on logarithmic scale, highlighting the divergence of the amplitude for larger values of $q$ as shown in appendix \ref{appendix:momamp}.}
    \label{fig:potential}
\end{figure}

 We see that in the cases introduces in this section, the potential serves as a confinement mechanism, taking the place of the system's size. The bi-scaling manifests as the transition of the FPT moments behavior, as seen in Fig.\,\ref{fig:potential}, from force-free to a potential-controlled dynamics. In the force-free regime, where the potential is negligible, the first-passage time statistics exhibit a well-known algebraic tail distribution with a persistence exponent $\theta=1/2$. However, as the time or the potential strength increases, a shift occurs towards the confined behavior, and the first-passage time distribution undergoes a transition characterized by the infinite density function.

\subsection{General force field}
In this section, we return to the general problem of FPT in a confining potential. Consider the dynamics of a one-dimensional Brownian particle influenced by an attracting potential field \(V(x)\). The Fokker-Planck equation for the propagator \(P(x,t)\) is given by Eq.\,(\ref{eq:FPeq2}). Physically free diffusion is expected when the temperature is high. We now consider this limit, but first we aim to rewrite this equation in dimensionless units. Doing so we identify the length, energy and time scale of the problem. Define the dimensionless variables $ y = x/L$ and $\tau = t/(L^2/D)$, where \(L\) is the characteristic confining length to be defined. Substituting these into Eq.\,(\ref{eq:FPeq2}) we get 
\begin{equation}\label{eq:fp12}
\frac{\partial }{\partial \tau} P(y,\tau)=  \left[ \frac{\partial^2 }{\partial y^2} + \frac{1}{T}\frac{\partial}{\partial y} \left( \frac{\partial V(Ly)}{\partial y} \right) \right]P(y,\tau)\, . \end{equation}
We consider a large class of potentials which can be written in the large-$x$ limit as 
\begin{equation}\label{eq:potential}
    V(x)\sim \kappa_\beta |x|^\beta \, , \,\,\,\, \, \, \, \, \, \beta \geq 1.
\end{equation}
The characteristic length $L$ is defined via the standard deviation of the Boltzmann equilibrium state:
\begin{equation}
    L^2 \equiv \int_0^\infty x^2 e^{-V(x)/T} \, dx / Z \, ,
\end{equation}
where \(Z\) is the normalization factor $Z=\int_0^\infty e^{-V(x)/T}\, dx$. From Eq.\,(\ref{eq:potential}), in the high-temperature limit, this evaluates to:
\begin{equation}\label{eq:forcelen}
L = \left(\frac{T}{\kappa_\beta}\right)^{1/\beta} \sqrt{\frac{\Gamma(3/\beta)}{\Gamma(1/\beta)}}\, .
\end{equation}
Using the asymptotic form of the potential Eq.\,(\ref{eq:potential}) and the confining length $L$ given in Eq.\,(\ref{eq:forcelen}) we can define $\tilde{V}(y)\equiv(\tilde{\kappa_\beta}/\kappa_\beta )  V(y)\sim \tilde{\kappa_\beta} y^\beta $ with  $\tilde{\kappa_\beta} = (\Gamma(1/\beta)/\Gamma(3/\beta))^{\beta/2}$. Eq.\,(\ref{eq:fp12}) is now
\begin{equation}\label{eq:fp13}
\frac{\partial }{\partial \tau} P(y,\tau)=  \left[ \frac{\partial^2 }{\partial y^2} + \frac{\partial}{\partial y} \left( \frac{\partial \tilde{V}(y)}{\partial y} \right) \right]P(y,\tau)\, . \end{equation} \par The eigenfunction equation corresponding to the Fokker-Planck Eq.\,(\ref{eq:fp13}) can then be written as
\begin{equation}
-\tilde{E}_n \phi_n(y) =  \left[ \frac{\partial^2 \phi_n(y)}{\partial y^2} + \frac{\partial}{\partial y} \left( \frac{\partial \tilde{V}(y) \phi_n(y)}{\partial y} \right) \right]\, ,  \end{equation}
where \(\phi_n(\cdot)\) and \(\tilde{E_n}\) are the eigenfunctions and the eigenvalues of the Fokker-Planck operator, respectively. The propagator \(P(y,\tau)\) can be generally written using the eigenfunction expansion
\begin{equation}\label{eq:generaleigfun}
    P(y,\tau)=\sum_n c_n \phi_n(y) e^{-\tilde{E}_n \tau}.
\end{equation}
Using the transformation \(\psi_n(y) = e^{V(y)/2} \phi_n(y)\) \cite{risken1996fokker}, we can rewrite Eq.\,(\ref{eq:generaleigfun}) as
\begin{equation}\label{eq:pb}
    P(y,\tau)=e^{\tilde{V}(y_0)/2-\tilde{V}(y)/2}\sum_n \psi_n(y_0) \psi_n(y) e^{-\tilde{E}_n \tau}\, ,
\end{equation}
where $P(y=0,\tau)=0$. The FPT PDF is defined by the current at the absorbing point, namely 
\begin{equation}
    \eta(\tau) = -\partial_y P(y,\tau)|_{y=0}\, ,
\end{equation}
which can be written as
\begin{equation}
    \eta(\tau)= e^{\tilde{V}(y_0)/2}\sum_n \psi_n(y_0) \partial_y \left(e^{-\tilde{V}(y)/2}\psi(y)\right)|_{y=0} \, e^{-\tilde{E}_n \tau}.
\end{equation}
Using the definition of the infinite density function $\mathcal{I}(\tau)$ as given in Eq.\.(\ref{eq:infdendefforce}), in the limit $y_0\ll 1$, we expand $\psi_n(y_0)\sim \psi'_n(0)y_0$ and $e^{\tilde{V}(y_0)/2}\sim 1$, obtaining
\begin{equation}\label{eq:etafin}
    \eta(\tau)\sim y_0\sum_n  |\partial_y \phi(y)|_{y=0}|^2 e^{-\tilde{E}_n \tau}.
\end{equation}
This formulation provides an explicit solution of the FPT PDF in the large-$L$ limit, from which we can extract the infinite density function as defined in Eq.\.(\ref{eq:infdendefforce}) to be
\begin{equation}\label{eq:etafinffun}
    \boxed{\mathcal{I}(\tau)=\sum_n  |\partial_y \phi(y)|_{y=0}|^2 e^{-\tilde{E}_n \tau}.}
\end{equation}
To show that this solution is non-normalizable, namely that in general to calculate the infinite density we need to use WKB approximation. More specifically, we will show that this solution diverges on the origin as $\mathcal{I}(\tau)\propto \tau^{-1+\theta}$ with $\theta=1/2$. Note that Eq.\,(\ref{eq:etafinffun}) is independent of the initial condition $r_0$, and in this sense universal. Further, this formula can be used to derive Eqs.\,(\ref{eq:infdenlin},\ref{eq:infho}) for particle in a constant force and harmonic potential. 

\subsubsection{WKB Approximation}
In the preceding section, we addressed the non-self-adjoint nature of the Fokker-Planck equation by employing a similarity transformation. As mentioned above, we used the transformation \(\psi_n(y) = e^{V(y)/2} \phi_n(y)\) \cite{risken1996fokker}, which allowed us to reduce the Fokker-Planck equation (\ref{eq:FPop}) to a Schrödinger-type equation with the same eigenvalues given by
\begin{equation}\label{eq:shrod} \left[-\frac{\partial^2}{\partial y^2} + U_{\textit{eff}}(y) \right] \psi_n(y) = \tilde{E}_n \psi_n(y), \end{equation}
where the effective potential \( U_{\textit{eff}}(y) \) is defined as
\begin{equation} 
 U_{\textit{eff}}(y) = \frac{1}{4} \left( \frac{\partial \tilde{V}(y)}{\partial y} \right)^2 - \frac{1}{2} \frac{\partial^2\tilde{V}(y)}{\partial y^2}. \end{equation}
For the general form of the potential given in Eq.\,(\ref{eq:potential}), the effective potential simplifies to:
\begin{equation}\label{eq:effpotantial}
U_{\textit{eff}}(y) \sim \frac{(\tilde{\kappa_\beta}\beta y^{\beta-1})^2}{4}\, ,
\end{equation}
for $y\gg 1$. The short-time limit of Eq.\,(\ref{eq:etafin}) is dominated by the higher eigenvalues \(\tilde{E}_n\). The WKB approximation provides a method for analyzing the system in the semiclassical limit, where it is the high-energy eigenstates that most contribute to the overall dynamics.
We start by evaluating the spectrum $\tilde{E}_n$, which in the semiclassical limit is determined by the Bohr-Sommerfeld quantization condition:
\begin{equation}\label{eq:spectrumdef} 
\int_{y_1}^{y_2} \sqrt{\tilde{E}_n - U_{\textit{eff}}(y)} \, dy = \left(n + \frac{3}{4}\right) \pi. \end{equation}
Here the $3/4$ comes from the absorbing boundary at the origin \cite{risken1996fokker}, and \(y_1\) and \(y_2\) being the classical turning points where \( \tilde{E}_n = U_{\textit{eff}}(y_{1,2}) \). Substituting the effective potential Eq.\,(\ref{eq:effpotantial}) into Eq.\,(\ref{eq:spectrumdef}), we obtain:
\begin{equation}
\pi(n+\frac{3}{4}) = \int_0^{y_{tp}} dy \sqrt{\tilde{E}_n - \frac{(\tilde{\kappa_\beta}\beta y^{\beta-1})^2}{4}},
\end{equation}
where \(y_{tp} = (4 E/\beta^2 \tilde{\kappa_\beta}^2)^{1/(2\beta-2)}\). By changing variables to \(z = y/y_{tp}\),  the integral becomes
\begin{equation}\label{densitystate}
\pi(n+\frac{3}{4}) = \sqrt{E} y_{tp} \int_0^1 dz \sqrt{1 - z^{2(\beta-1)}}=\left( \frac{2}{\beta \tilde{\kappa_\beta}} \right)^{1/(\beta-1)} E^{\beta/(2\beta-2)} \frac{\sqrt{\pi}\Gamma(\frac{1}{2\beta-2})}{2\beta\Gamma(\frac{\beta}{2\beta-2})}.
\end{equation}
In the WKB approximation, the eigenfunctions \(\psi_n(\cdot)\) in Eq.\,(\ref{eq:shrod}) are approximated as:
\begin{equation}\label{eq:psi1}
\psi_n(y) \approx \frac{A}{\sqrt{k(y)}} \exp \left( \pm i \int_0^y \sqrt{\tilde{E}_n - U_{\textit{eff}}(y')} dy' \right)\sim \frac{\sin{\left( \int_0^y dy \sqrt{E - U_{eff}(y')} \right)}}{(\tilde{E}_n - U_{eff}(y))^{1/4}}\, .\end{equation}
In the short-time limit, corresponding to the small $y$ limit, we take \(y \sim 0\), so Eq.\,(\ref{eq:psi1}) simplifies to:
\begin{equation}
\psi_n(y) \sim \frac{\sin{(\sqrt{\tilde{E}} y)}}{\tilde{E}^{1/4}} = y\,\tilde{E}^{1/4}\, .
\end{equation}
For normalization, considering the fast oscillations and that contributions beyond \(y_{tp}\) are negligible, we find:
\begin{equation}
\int_0^\infty dy |\psi(y)|^2 \sim \int_0^\infty dy \frac{1/2}{(\tilde{E} - u_{eff}(y))^{1/2}} = \frac{1}{2} \frac{y_{tp}}{\sqrt{\tilde{E}}} \int_0^1 dz (1 - z^{2(\beta-1)})^{-1/2}
\end{equation}
Thus, the integral evaluates to:
\begin{equation}
\int_0^\infty dy |\psi(y)|^2 \sim \frac{y_{tp}}{\sqrt{\tilde{E}}} \frac{\sqrt{\pi}\Gamma(\frac{1}{2\beta-2})}{2(2-\beta)\Gamma(-1+\frac{\beta}{2\beta-2})}.
\end{equation}
Finally, from Eq.\,(\ref{eq:etafin}), \(\eta(t)\) is expressed as:
\begin{equation}
\eta(\tau) \sim y_0 \sum_n \sqrt{\tilde{E}}\frac{1}{\int_0^\infty dy |\psi(y)|^2} e^{-\Tilde{E} \tau}\, .
\end{equation}
In the high-energy limit, the spacing between energy levels becomes small relative to the typical energy, meaning \(\tilde{E}_n \gg \Delta E\). This allows us to approximate the sum over \(n\) as an integral. Specifically, we can replace \(\sum_n\) with \(\int dE \frac{dn}{dE}\), where \(\frac{dn}{dE}\) is the density of states calculated from Eq.\,(\ref{densitystate}). Recovering the units of $\eta(t)$, this leads to:
\begin{equation}\label{eq120}
\eta(t) \sim \frac{x_0}{\pi L^3} \int_0^\infty d\Tilde{E} \sqrt{\Tilde{E}} e^{-\Tilde{E} t/L^2}\sim \frac{x_0}{\sqrt{4 \pi}} t^{-3/2}.
\end{equation}
To summarise, for any potential field of the form given in Eq.\,(\ref{eq:potential}), Eq.\,(\ref{eq120}) gives $\mathcal{I}(\tau)\propto \tau^{-3/2}$ for $\tau  \rightarrow 0$. This implies that $\mathcal{I}(\cdot)$ in Eq.\,(\ref{eq:etafinffun}) is not normalizable, still as throughout this work it gives the moments of order $q>\theta$ ($\theta =1/2$) of the process, including the first moment (see appendix \ref{appendix:wkbmom}). This shows the wide applicability of the infinite density concept, as it holds generally valid for a very general class of attractive force fields.

\section{Stochastic resetting}
The principles detailed in this manuscript can be extended to a model of stochastic resetting for a diffusive particle, where the particle's position is reset randomly over time at a rate \(r\) \cite{bhat2016stochastic,gupta2022stochastic,evans2011diffusion,yin2023restart,ray2020diffusion,dubey2023first,singh2022first}. This resetting process corresponds to Poissonian resetting, returning the particle to a fixed point $x_r$. In addition to the resetting we add a target at $x=0$, and focus on the first passage time PDF. In the short-time limit, we expect the FPT distribution to behave like a free particle, namely the first scaling function $f_{\infty}(t)$ is given by Eq.\,(\ref{eq:freesol}). This short-time regime is characterized by the particle's initial diffusion before the effects of resetting become significant. In the long-time limit, the resetting process leads to a cutoff in the distribution. We consider the limit where both the FPT \(t\) and \(1/r\) are large, but the product \(r\,t\) remains constant, from which we obtain the infinite density function. We consider the case of free diffusion between resetting events, in the absence of a reflecting wall or force field, and will later define the length scale \(L\) within this context.
\par
\begin{figure}[h]
    \centering
    \includegraphics[width=1\textwidth]{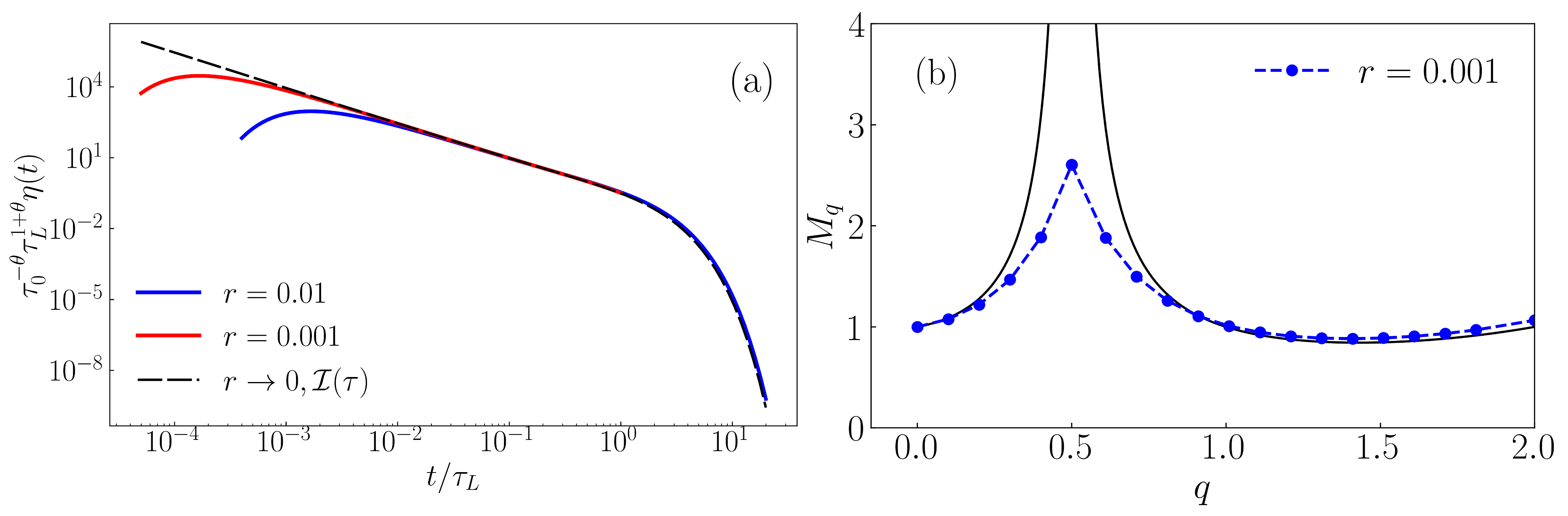}
    \caption{ (a) The rescaled function $\tau_L^{3/2}\eta(t)$ for the resetting problem, given by the inverse Laplace transform of Eq.\,(\ref{eq:laplacereset}), converges to the infinite density Eq.\,(\ref{eq:idreset}). (b) The asymptotic behavior of the FPT moments where $q_c=\theta=1/2$.  Here we took $D=1$ and $x_0=1$.}
    \label{fig:resetfig}
\end{figure}

 Consider a resetting process in a semi-infinite domain $x\in[0,\infty)$, where the fixed point is located at the initial position, namely $x_r=x_0$. The Laplace transform $t\rightarrow s$ of the survival probability is given by \cite{evans2020stochastic}
\begin{equation}\label{eq:resetsurv}
    \hat{S}(s)=\frac{1-\exp(-x_0\sqrt{\frac{s+r}{D}})}{s+r\,\exp(-x_0\sqrt{\frac{s+r}{D}})}\, ,
\end{equation}
where $D$ is the diffusion constant describing the Brownian motion between resetting. The FPT PDF is defined as 
\begin{eqnarray}
    \hat{\eta}(s)=1-s\hat{S}(s).
\end{eqnarray}
Substituting $\hat{S}(s)$ from Eq.\,(\ref{eq:resetsurv}), we get
\begin{equation}\label{eq:laplacereset}
    \hat{\eta}(s)=\frac{(r+s)\exp(-x_0\sqrt{\frac{s+r}{D}})}{r\exp(-x_0\sqrt{\frac{s+r}{D}})+s}.
\end{equation}
Consider the long-time limit, where $x_0^2/D\ll 1/s$, and then Eq.\,(\ref{eq:laplacereset}) can be approximated as
\begin{equation}\label{eq:laplaceresetapprox}
   \hat{\eta}(s) \simeq 1-\frac{x_0\sqrt{r}}{\sqrt{D}}\frac{(s/r)}{\sqrt{1+(s/r)}}.
\end{equation}
The inverse Laplace transform of Eq.\,(\ref{eq:laplaceresetapprox}) is given by
\begin{equation}
   \eta(t) \sim \frac{x_0 r^{3/2}}{\sqrt{\pi D}}e^{-rt}\left((rt)^{-3/2}/2+(rt)^{-1/2} \right).
\end{equation}
The typical time-scale corresponding to the long-time limit of this problem involves the resetting rate $\tau_L=1/r$. Note that in this problem the diverging length scale is defined by $L=\sqrt{ D/r}$ as we are considering the rare resetting limit. In previous problems, the system's size or the potential's strength dictated the cutoff behavior. Here the resetting rate $r$ acts as the confining mechanism. From Eq.\,(\ref{eq:infdefgeneral}), the infinite density function is defined as
\begin{equation}\label{eq:reslim}
   \mathcal{I}(\tau) \equiv\lim_{t,1/r \rightarrow \infty}\text{ }  r^{-3/2} (x_0/\sqrt{D})^{-1}\eta(t)\, ,
\end{equation}
where $\tau=r\,t$ and the persistence exponent is $\theta=1/2$, gives
\begin{equation}\label{eq:idreset}
  \boxed{ \mathcal{I}(\tau) =\frac{e^{-\tau}}{\sqrt{\pi}}\left(\tau^{-3/2}/2+\tau^{-1/2} \right).}
\end{equation}
For small $\tau$, $\mathcal{I}(\tau) \propto \tau^{- 3/2}$, and so $\mathcal{I}(\tau)$ is indeed non-normalizable. In panel (a) of Fig.\,\ref{fig:resetfig} we see that the intermediate to long-time limit of the distribution converges to Eq.\,(\ref{eq:idreset}). 
\par In the limit of small rate, namely $1/r\gg x_0^2/D$, the moments are given by
\begin{equation}\label{eq:asymp_momrese}
    \langle t^q\rangle\xrightarrow{}\begin{cases}
                     \tau_0^q M_q^{-}=\langle t^q\rangle_{\infty} , &  q<1/2\\
                     r^{1/2-q}\tau_0^{1/2} M_q^{+}=r^{1/2-q} \tau_0^{1/2}\int_0^\infty \tau^q \mathcal{I}(\tau)d\tau , &  q>1/2\\
                    \end{cases} \, ,
\end{equation}
where $\tau_0=x_0^2/D$. The low-order moments $\langle t^q\rangle_{\infty}$ are the FPT moments of a free particle as given in Eq.\,(\ref{asymp0}). The higher-order moments $q>1/2$ are given by the moments calculated from Eq.\,(\ref{eq:idreset})
\begin{equation}\label{eq:highmomres}
    \int_0^{\infty}d\tau \mathcal{I}(\tau)t^q = q\Gamma(q-1/2)/\sqrt{\pi}.
\end{equation}
In panel (b) of Fig.\,\ref{fig:resetfig} we plot the dimensionless $M^\pm_q$ vs. $q$. The moments were calculated by numerical integration of the inverse transform of Eq.\,(\ref{eq:laplacereset}). The $q<1/2$ moments converge to those of free diffusion since these moments correspond to the times when the system has not yet reset. These moments clearly do not depend on the rate $r$.

\section{Run and tumble particle under resetting}
Next we study the dynamics of a run-and-tumble particle (RTP) under stochastic resetting. The RTP model consists of alternating phases of straight-line motion (run phase) at a constant speed \(v_0\) and random reorientation (tumble phase), resulting in a sudden change in direction \cite{lovely1975statistical,berg1972chemotaxis}. The tumble events occur instantaneously at a constant rate \(\gamma\).

We consider an RTP starting at the absorbing point \(x_0 = 0\), with an initial velocity of either \( \pm v_0 \), each with a probability of \(1/2\). The particle is reset to \(x = 0\) at a constant rate \(r\), with the velocity also reset to either \( \pm v_0 \). In the limit of a slow resetting rate \(r \rightarrow 0\), the system effectively behaves as a free RTP, meaning the FPT distribution converges to that of an RTP particle without resetting, given by 
\cite{basu2023target,mori2020universal}
\begin{equation}\label{eq:rntfinf1}
    f_{\infty}(t)= \frac{1}{2t} e^{-\gamma t} I_1(\gamma t)  \, ,
\end{equation}
where $I_1(\cdot)$ is the modified Bessel function of order $1$. In the limit of large \(t\) (i.e., \( t \gg 1/\gamma \)), using the expansion of the modified Bessel function $I_1(x)\sim e^x/\sqrt{2\pi x}$, Eq.\,(\ref{eq:rntfinf1}) simplifies to 
\begin{equation}\label{eq:rntfinf}
    f_{\infty}(t)\sim   \frac{1}{2\sqrt{2\pi \gamma }}t^{-3/2}  \, ,
\end{equation}
 Thus, from Eq.\,(\ref{eq:rntfinf}) we see that the the persistence exponent is $\theta=1/2$.
 
\par For a RTP with resetting, the Laplace transform of the FPT distribution was derived by Evans and Majumdar using a renewal approach \cite{evans2018run}. The exact expression is given by:
\begin{equation}\label{eq:exactrnt}
    \hat{\eta}(s)=1-\frac{s}{r}\left( -1+\frac{2\gamma(s+r)}{2\gamma s-r(\sqrt{(r+s)(r+s+\gamma)}-(r+s+2\gamma)} \right).
\end{equation}
Consider the limit of slow resetting, where the resetting rate is much slower compared to the tumble rate, i.e., \( 1/r \gg 1/\gamma \), and long times, where \( 1/s \gg 1/\gamma \), while keeping \( r/s \) constant. In this limit, Eq.\,(\ref{eq:exactrnt}) simplifies to
\begin{equation}\label{eq:asymrntlap}
   \hat{\eta}(s) \sim 1-\frac{s/r}{\sqrt{2\gamma(+s/r)}}\, ,
\end{equation}
and the inverse transform of Eq.\,(\ref{eq:asymrntlap}) is given by
\begin{equation}
    \eta(t)\sim\frac{r^{3/2}}{\sqrt{2\gamma \pi}}e^{-r t}\left(\frac{1}{2}(r\,t)^{-3/2}+(r\,t)^{-1/2}\right).
\end{equation}
The first time scale, associated with the tumble rate, is \(\tau_0 = (2\gamma)^{-1}\), while the second time scale, corresponding to the resetting rate, is defined as \(\tau_L = 1/r\). Following that, and from Eq.\,(\ref{eq:infdefgeneral}), the infinite density function is defined as
\begin{equation}\label{eq:reslim}
   \mathcal{I}(\tau) \equiv\lim_{t,1/r \rightarrow \infty}\text{ }  \sqrt{2\gamma}\,r^{-3/2} \eta(t)\, ,
\end{equation}
where $\tau=r\,t$, we find
\begin{equation}\label{eq:idrnt}
   \mathcal{I}(\tau) =\frac{e^{-\tau}}{\sqrt{4\pi}}\left(\tau^{-3/2}+2\,\tau^{-1/2} \right).
\end{equation}
The scaling function in Eq.\,(\ref{eq:idrnt}) is identical to the one derived in Eq.\,(\ref{eq:idreset}). This is expected, as the long-time limit of Eq.\,(\ref{eq:rntfinf}) shows that the RTP dynamics reduce to those of ordinary Brownian motion with an effective diffusion coefficient $D_{\textit{eff}}=v_0^2/2\gamma$. Since the particle starts at the origin, namely $x_0=0$, we define an effective length scale $x_0^{\textit{eff}} \equiv v_0/2\gamma$, the characteristic distance the particle is expected to move away from the target before the first tumble occurs. Following this definition, we can define the first time scale straightforwardly as $\tau_0\equiv(x_0^{\textit{eff}})^2/D_{\textit{eff}}=1/2\gamma$. As a result, the scaling function corresponding to the long-time limit, given in Eq.\,(\ref{eq:idrnt}), mirrors the one found for stochastic resetting of a Brownian particle in previous section, with the time scale \(\tau_0\) is now relate to the tumble rate of the system.
\par
\begin{figure}[h]
    \centering
    \includegraphics[width=1\textwidth]{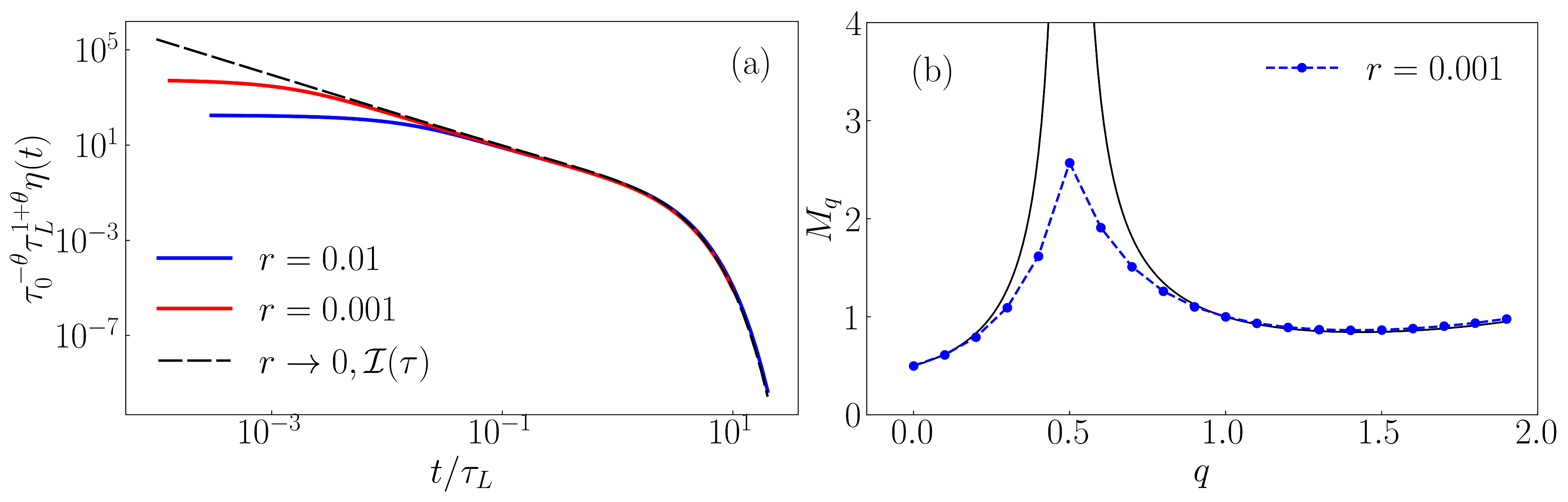}
    \caption{ (a) The rescaled function $\tau_L^{3/2}\tau_0^{-1/2}\eta(t)$ for run and tumble particle under resetting, given by the inverse Laplace transform of Eq.\,(\ref{eq:exactrnt}), converges to the infinite density Eq.\,(\ref{eq:idrnt}). (b) The asymptotic behavior of the FPT moments where $q_c=\theta=1/2$.  Here we took $\gamma=2$ and $v_0=1$.}
    \label{fig:rntfig}
\end{figure}

 In panel (a) of Fig.\,(\ref{fig:rntfig}) we show the converges of the rescaled FPT distribution, given by the inverse Laplace transform of Eq.\,(\ref{eq:exactrnt}) to the scaling function $\mathcal{I}(\tau)$ as we increase $\tau_L=1/r$.  

\par In the limit $1/r\gg 1/2\gamma$, namely the tumble rate is much higher than the resetting rate, the FPT moments can be expressed as
\begin{equation}\label{eq:asymp_momrnt}
    \langle t^q\rangle\xrightarrow{}\begin{cases}
                     (1/2\gamma)^q M_q^{-}=\langle t^q\rangle_{\infty} , &  q<1/2\\
                     r^{1/2-q}(1/2\gamma)^{1/2} M_q^{+}=r^{1/2-q} (1/2\gamma)^{1/2}\int_0^\infty \tau^q \mathcal{I}(\tau)d\tau , &  q>1/2\\
                    \end{cases} \, .
\end{equation}
The low-order moments for $q<1/2$ are the FPT moments of the RTP without resetting in Eq.\,(\ref{eq:rntfinf1}), given by
\begin{equation}
    M_q^- = \Gamma(1+q)\,_2F_1\left(\frac{1+q}{2},\frac{2+q}{2},2,1\right)\, .
\end{equation}
The higher-order moments $q>1/2$ are given by the moments calculated from Eq.\,(\ref{eq:idreset}), given explicitly in Eq.\,(\ref{eq:highmomres}). In panel (b) of Fig.\,\ref{fig:rntfig} we plot the dimensionless $M^\pm_q$ vs. $q$. These moments  are computed through the numerical integration of the inverse Laplace transform of Eq.\,(\ref{eq:exactrnt}). We see that the $q<1/2$ moments converge to those of the RTP dynamics without resetting, and therefore do not depend on the resetting rate $r$. In contrast, the $q>1/2$ moments align with the behavior of the resetting problem without RTP dynamics. Therefore the only dependence on the tumbling results from the scaling time $\tau_0=1/2\gamma$.

 \section{Uniform approximation}\label{sec:uniform}
In this section, we discuss the uniform approximation for the FPT statistics. The FPT PDF is characterized by two scaling functions, namely, $f_{\infty}(t)$ and $\mathcal{I}(t)$, which describe its asymptotic behavior for short and long times, respectively, as defined in Section \ref{sec:scaling}. As we showed, there is a region in time $t$, that the two solutions match, and hence we can construct a uniform approximation that smoothly connects the two approximations. 
The main idea behind the method is to assume that the density function varies slowly compared to the characteristic scales of the system, allowing us to obtain approximate solutions that capture a wide range of time scales. As we found in section \ref{sec:scaling}, the uniform approximation of the FPT PDF is given by
\begin{equation}\label{eq:uniform2}
     \eta(t)\sim \mathcal{N}\tau_0^{-1}f_{\infty}(t/\tau_0)\tau_L^{-1-\theta}\mathcal{I}(t/\tau_L)t^{1+\theta} \, .
\end{equation}
For each of the models presented in this work, we provided an explicit expression for both $f_{\infty}(t)$ and $\mathcal{I}(t)$. For example for the Harmonic field, we use $\mathcal{I}(\tau)$ given in Eq. (\ref{eq:infho}) and $f_{\infty}(t)$ is the PDF for a force-free particle given in Eq. (\ref{eq:freesol}), and find $\eta(t)\sim 4\sqrt{2} x_0 D\kappa^{3/2} e^{-D\kappa t-x_0^2/4Dt}(1-e^{-2 D\kappa t})^{-3/2}/\sqrt{\pi}$.
Eq. (\ref{eq:uniform2}) provides good approximation for both short and long time-scales. In Table II we list the uniform approximations found for the models under study.

\begin{figure}[h!]
    \centering
    \includegraphics[width=0.7\textwidth]{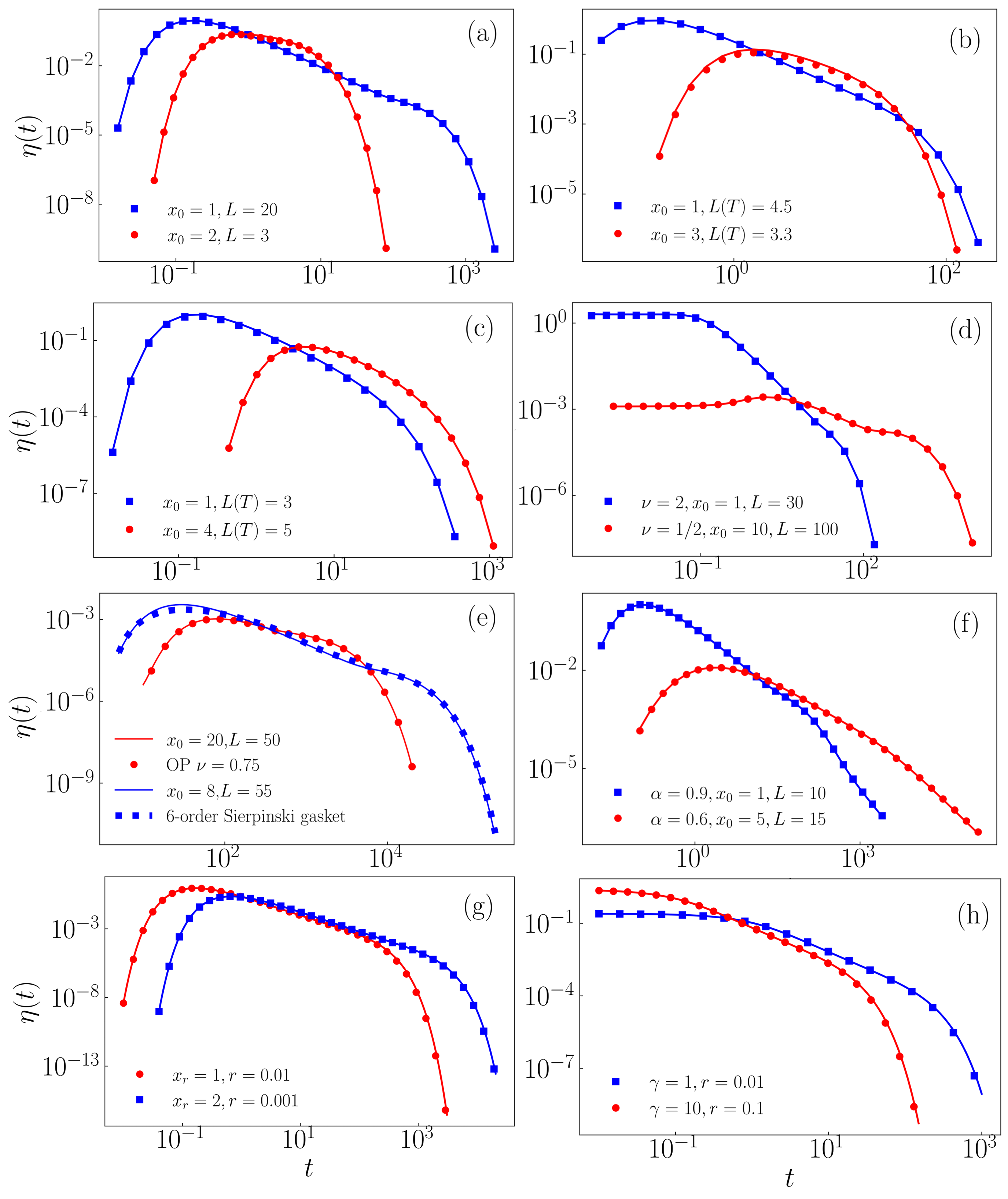}

    \caption{FPT PDF: Uniform approximation (solid lines), described by the expressions in Table \ref{table:2}, is compared with the exact solution (symbols) for various models:(a) 1D box, (b) V-shaped potential, (c) Harmonic oscillator, (d) 2D wedge, (e) fractal geometry, (f) CTRW, (g) Stochastic resetting, and (h) Run and tumble with resetting. }
    \label{fig:unifrom}
\end{figure}

 \par For the case of diffusion in two-dimensional wedge (Fig. \ref{fig:unifrom} panel (b)), the short-time limit of the PDF does not decay to zero as is typical for many other systems. Instead, it converges to a constant value. This behavior arises due to the particular chosen initial condition of the angular part which allows the particle to start infinitesimally close to the absorbing line, hence the FPT is zero, and the probability of reaching the target in extremely short times remains non-negligible.
\par In certain experimental scenarios, a more realistic initial condition may not be precisely specified by a single value, but rather by a distribution. For instance, the initial condition could be represented by a narrow Gaussian distribution centered around $x_0$. In such cases, the resulting distribution can be calculated by integrating over the initial condition distribution, given by the expression $\int_0^L dx_0 \, p(x_0) \, \eta_{x_0}(t)$, where $\eta_{x_0}(t)$ represents the FPT PDF and $p(x_0)$ represents the distribution of the initial condition. 
\begin{table}[h]
\centering
\begin{tabular}{||c c c c c c c c||} 
 \hline
 model & $d_f$ & $d_w$ & $\theta$ &$q_c$& $\tau_0$& $\tau_L$& $\Tilde{K}$ \\ [0.5ex] 
 \hline\hline
 1d box & 1 & 2 & 1/2 &1/2&$x_0^2/D$ &$L^2/D$& $2D$\\ 
 wedge & 2 & 2 & $\nu/2$  &$\nu/2$ &$r_0^2/D$&$L^2/D$&$4D$\\
 harmonic potential & 1 & 2 & 1/2 &1/2 &$x_0^2/D$&$(\kappa D)^{-1}$ &$2D$\\
 V-shaped potential & 1 & 2 & 1/2 & 1/2&$x_0^2/D$& $(F^2 D)^{-1}$&$2D$\\
 CTRW & 1 & $2/\alpha$ & $\alpha/2$  &$\alpha/2$ &$(x_0^2/K_{\alpha})^{1/\alpha}$& $(L^2/K_{\alpha})^{1/\alpha}$&$2K_{\alpha}/\Gamma(1+\alpha)$\\
  fractal & $d_f$ & $d_w$ & $1-d_f/d_w $  &$1-d_f/d_w$ &$r_0^{d_w}/K$& $L^{d_w}/K$&$\frac{\Gamma((2+d_f)/d_w)}{\Gamma(d_f/d_w)}(d_w^2 K)^{2/d_w}$\\
  resetting &$1$&$2$&$1/2$&$1/2$&$x_0^2/D$&$1/r$&$2D$\\ 
  run and tumble with resetting &$1$&$2$&$1/2$&$1/2$&$1/(2\gamma)$&$1/r$&$v_0^2/(2\gamma)$\\ [1ex] 
 \hline
\end{tabular}
\caption{Scaling parameters for the different models.}
\label{table:1}
\end{table}

\begin{table}[h]
\centering
\renewcommand{\arraystretch}{2.5}
\begin{tabular}{||c c||} 
 \hline
 model  &uniform approximation for $\eta(t)$ \\ [0.5ex] 
 \hline\hline
 1d box  &  $-\frac{x_0}{L}e^{-x_0^2/4Dt}\partial_{t }\vartheta_2\left(e^{-\pi^2 D t/L^2}\right)$ \\ 
 
 wedge & $ r_0L^{-2-\nu}D^{\frac{1+\nu}{2}}\frac{\Gamma(\frac{3+\nu}{2})}{2^{1-nu}\Gamma(2+\nu)}e^{-\frac{r_0^2}{8 D t}}t^{\frac{\nu-1}{2}}\left( I_{\frac{3+\nu}{2}}(\frac{r_0^2}{8 D t}) - (1-\frac{8Dt}{r_0^2}(1+\nu))I_{\frac{1+\nu}{2}}(\frac{r_0^2}{8 D t}) \right)\sum_k c_k k^{2+\nu}e^{-\frac{Dk^2}{L^2}t}$\\
 
  CTRW & $-2\Gamma(-\frac{\alpha}{2})L^{-3}x_0^{-2/\alpha}t^{3\alpha/2}\ell_{\alpha/2}\left(\frac{t}{x_0^{2/\alpha}}\right) \sum_n \pi^2(n+\frac{1}{2})^2 E_{\alpha,\alpha}\left(-\pi^2(n+\frac{1}{2})^2 t^{\alpha}/L^2\right)$\\

  fractal & $ K r_0^{d_w-d_f}L^{-2d_w+d_f}\frac{d_w 2^{2\nu-3} \Gamma(\nu) }{ \Gamma(2-\nu)}e^{-\frac{r_0^{d_w}}{d_w^2 K t}} \sum_{n=0}^{\infty}  \frac{J_{\nu}(z_{-\nu,n}) }{J_{1-\nu}(z_{-\nu,n})}  z_{-\nu,n}^{3-2\nu}  e^{-d_w^2 z_{-\nu,n}^2K t/4L^{d_w}}$\\ 

   V-shaped potential &$\frac{x_0}{\sqrt{{4 D\pi}}}t^{-3/2}e^{-(DF/k_B T)^2 t/4-x_0^2/4Dt}$\\
  
 Harmonic potential &$x_0 D\kappa^{3/2}\sqrt{\frac{2}{\pi}} \frac{4 e^{-D\kappa t-x_0^2/4Dt}}{(1-e^{-2 D\kappa t})^{3/2}}$ \\ 
  Stochastic resetting &$e^{-x_0^2/4Dt}\frac{x_0}{\sqrt{4\pi\, D}}e^{-r\,t}t^{-3/2}\left(1+2r\,t \right)$ \\ 
   Run and tumble with resetting &$r^{3/2}t^{1/2}e^{-rt-\gamma t}I_1(\gamma t)\left( (rt)^{-3/2}/2+(rt)^{-1/2} \right)$ \\

 \hline
\end{tabular}
\caption{Uniform approximation for each of the models discussed in this paper, obtained using Eq. (\ref{eq:uniform2}).}
\label{table:2}
\end{table}

\section{Discussion}
The FPT PDF $\eta(t)$ for compact search in large but finite systems with localized initial conditions consists of three parts: a singular phase at short times where the particle cannot reach the target, a peak at intermediate times, and a cutoff at long times. The distribution has the following properties:

\begin{itemize} 

\item[1.]
{\em Factorization:} 
The problem involves two length scales, the system size and the initial distance between the particle and the target, leading to two time scales, $\tau_L$ and $\tau_0$, correspondingly. Our bi-scaling theory demonstrates that the FPT statistics can be described as a product of two functions, $f_\infty(\cdot)$ and $\mathcal{I}(\cdot)$. This simplification allows for the separate evaluation of these functions. The factorization of the solution into a product of two functions has many consequences, which are discussed in the following points.

\item[2.] {\em Bi-Scaling.}  
The function $f_\infty(\cdot)$ represents the distribution of the first passage time for an infinite system, while the function $\mathcal{I}(\cdot)$, which is not normalized, describes another asymptotic regime arising from the finiteness of the system. The existence of a scaling range where $\tau_L \gg t \gg \tau_0$ guarantees the matching of the large-time limit of $f_\infty(\cdot)$ and the small-time limit of $\mathcal{I}(\cdot)$ match, both relating to the persistence exponent. 

\item[3.] {\em Two Limits of the PDF of First Passage Times.}
Mathematically, by taking the limit as $L \to \infty$ while keeping $t$ fixed, the PDF of first passage times, $\eta(t)$, yields $f_\infty(t)$. In the second limit, where both $t$ and $\tau_L$ are large but their ratio remains finite, we obtain $\mathcal{I}(\tau)$. These functions are essential tools in the theory of first passage times for compact search.

\item[4.] {\em Three Exponents:}
The problem is characterized by three exponents: the widely used walk dimension $d_w$ and the persistence exponent $\theta$, as well as a new exponent $\gamma$. The amplitudes of the moments $M_q$ diverge as $q$ approaches $\theta$ from above and below, following the relation $M_q \sim |q - \theta|^{-\gamma}$. Unlike $\theta$ and $d_w$, which can vary, $\gamma=1$ is universal across all models. Close to $q = \theta$, the convergence is extremely slow, as illustrated in the corresponding figures \ref{fig:1dbox}(b),\ref{fig:wedgedemons}(c),\ref{fig:ctrw}(b), etc.

\item[5.] {\em Observables:}
 In the context of infinite ergodic theory, the type of observable under study is crucial. The moments of the first passage time, $\langle t^q \rangle$, can be characterized as either integrable for $q > \theta$ or non-integrable for $q < \theta$ with respect to the non-normalized invariant density $\mathcal{I}(\cdot)$. These two classes of observables exhibit vastly different behaviors. The non-integrable low-order moments, $q < \theta$, are not sensitive to system size, as they describe the typical behavior of the PDF of first passage times. In contrast, the moments for $q > \theta$ provide information on the size of the domain. Notably, the first moment, the mean, does not hold special importance.

\item[6.] {\em Spectrum of Exponents:} 
The moments depend on the size of the system and the initial distance between the particle and the target, or more generally, the absorbing boundary condition. Two piecewise linear functions, $q \mu(q)$ and $q \nu(q)$, describe the scaling behavior, with a point of discontinuity at $q = \theta$. Measuring these functions provides both $\theta$ and $d_w$. Furthermore, the functions $\nu(q)$, describing the dependence of moments on the size, and $\mu(q)$, describing the dependence on the initial condition, obey a simple sum rule in Eq.\,(\ref{etaa}).

\item[7.] 
{\em General Formal Solution:}
The general formal solution of these types of problems is given by an infinite series eigenfunction expansion. This type of solution is generally non-separable and does not typically imply that factorization is possible. Additionally, this solution depends on the initial condition, the shape of the domain, and other factors. In contrast, the scaling function $\mathcal{I}(\tau)$, which was the main
focus of the technical part of this work, is independent of the initial condition and does not depend explicitly on the system size, as the latter only affects the rescaling of the first passage time, see for example Eq.\,(\ref{eq:inf1first}).

\item[8.] 
{\em Closed-Form Solutions for Special-Cases:}
The function $\mathcal{I}(\tau)$ has been found in closed forms for special cases, such as a particle in a box, in a harmonic field, and for the resetting problem, see Eqs.\,(\ref{eq:inf1first},\ref{eq:infho},\ref{eq:idreset}). These simple expressions are valuable and can be obtained using techniques like the Mellin transform or summation over infinite series. While there is no general master function that solves the problem for all cases, we have determined the functions $\mathcal{I}(\tau)$ for both small $\tau$ (where it diverges) and large $\tau$ for all models under study. The key advantage of $\mathcal{I}(\tau)$ is that its calculation does not involve the initial condition, making it mathematically easier to handle compared to the full solution. Moreover, $\mathcal{I}(\tau)$ can be evaluated from numerical and experimental data, requiring the exponents $d_w$ and $\theta$. The persistence exponent $\theta$ arises from the power-law tail of the first passage time PDF for an infinite system, while $d_w$ is related to free diffusion exponents or alternatively they can be  directly extracted  from the spectrum of exponents describing first passage time moments, as mentioned.

\item[9.] {\em Force Fields:}
 The theory presented here is valid for a particle in a confining force field and for systems with hard or absorbing walls. In the former case, the length scale $L$ is evaluated from the Boltzmann equilibrium, making it temperature-dependent. This highlights the generality of our theory, which should also apply to other systems, such as active systems. Using the WKB method and the Fokker-Planck equation, we showed that $\mathcal{I}(\tau)$ diverges as $\tau^{-3/2}$ for short times, which holds for a wide class of potentials that increase linearly or faster at large $x$. Weaker potentials fall into a different category.
The $\mathcal{I}(\tau) \propto\tau^{-3/2}$ is a special case of the more general behavior
$\mathcal{I}(\tau) \propto \tau^{ - 1- \theta}$.
 
\item[10.] {\em Non-Markovian Dynamics and Disordered systems.}
Our study primarily focused on Markovian dynamics, where the exact solution for $\eta(t)$ involves an infinite series of exponentially decaying functions. However, this characteristic does not apply to strongly disordered systems, which we modeled using semi-Markovian Continuous Time Random Walks (CTRW). In these cases, the sum of exponential decays is absent, with $\mathcal{I}(\cdot)$ diverging for short times and decaying as a power law at large times. 
We propose that the bi-scaling theory can extend to other recurrent diffusion models in disordered environments. We also believe the theory may apply to underdamped Langevin dynamics and systems with short-term memory described by the generalized Langevin equation. Given the presence of the exponent $\theta$ in fractional Brownian motion, we speculate that the theory may also be relevant for processes with long-term correlations, provided the search is compact and the system large.

\item[11.] {\em Non-equilibrium systems:} Our results show that bi-scaling applies to the resetting paradigm, which inherently represents a non-equilibrium steady state. In contrast, the model of diffusion in a force field assumes thermal equilibrium, but this condition is not required, as the bi-scaling theory developed here is broadly applicable.

\item[12.] {\em Generality and flexibility:}
The theory is highly general and applicable to different shapes of distributions. It covers models like CTRW, where the FPT distribution shows a power-law decay, and it also handles unique cases like the 2D wedge model, where the first scaling function remains constant instead of decreasing to zero when $t\rightarrow 0$, see Fig.\,\ref{fig:wedgedemons}. The theory can be applied without exact knowledge of the analytical form of the scaling functions. For instance, in the case of the Sierpinski gasket, the second scaling function is determined using the Shaughnessy-Procaccia model, while the first scaling function does not have an explicit analytical solution. The theory remains effective even though one of the scaling functions only has a numerical form. In this model, we observe the matching of the two scaling functions, even though the first one describes a discrete space.

\end{itemize}

\section{Acknowledgments}
We would like to thank to the Israel Science Foundation for their support through grant 1614/21. Additionally, we extend our sincere thanks to Olivier Bénichou and Léo Régnier for valuable discussions and insights.

\newpage
\appendix
\section*{Appendices}
\section{Particle in a box - calculation of the infinite density using the Mellin transform}\label{appendix_mom1a}

The first-passage PDF in Laplace space, denoted as $\hat{\eta}(s)$, is expressed as \cite{redner2001guide}:
\begin{equation}
    \hat{\eta}(s) =  \frac{\cosh\left(\sqrt{\frac{s}{D}}(x_0-L)\right)}{\cosh\left(\sqrt{\frac{s}{D}}L\right)},
\end{equation}
which can be further simplified to:
\begin{equation}\label{eq:appendixlaplace1d}
    =  \cosh\left(\sqrt{\frac{s}{D}}x_0\right)-\sinh\left(\sqrt{\frac{s}{D}}x_0\right)\tanh\left(\sqrt{\frac{s}{D}}L\right).
\end{equation}
In the limit $s\tau_0/\tau_L\ll 1$ \cite{benichou2014first}, we have the following approximation:
\begin{equation}\label{eq:lapexpand_1d}
    \hat{\eta}(s) \approx 1-\frac{x_0}{L}\sqrt{\Tilde{s}}\tanh\left(\sqrt{\Tilde{s}}\right),
\end{equation}
which can also be written as:
\begin{equation}\label{append_mom1}
    = 1-2\frac{x_0}{L}\sum_{m=0}^\infty \frac{\hat{s}}{\pi^2(m+\frac{1}{2})^2+\hat{s}},
\end{equation}
where $\Tilde{s} = sL^2/D$. By expanding the PDF $\hat{\eta}(s)$ using the FPT moments, defined as $\hat{\eta}(s) = \sum_{n=0}^\infty(-1)^n \frac{s^n}{n!}\langle t^n\rangle$, we obtain the expression for the $n$th moment as:
\begin{equation}\label{eq:exmom1}
    \langle t^n\rangle_{n\in \mathbb{Z_+}} \sim 
                     2\tau_L^{n-\frac{1}{2}}\sqrt{\tau_0}\text{ }n! \underbrace{\sum_{m=0}^{\infty} \left(\pi(m+\frac{1}{2})\right)^{-2n}}_{\pi^{-2n}(2^{2n}-1)\zeta(2n)} ,
\end{equation}
where $\zeta(.)$ is the Riemann zeta function, $\tau_L=L^2/D$ and $\tau_0=x_0^2/D$. Clearly from normalization when $n=0$, $\langle t^0\rangle =1$. Eq.\,(\ref{eq:exmom1}) is valid for any positive integer.
 If  we assign a real value $n$, we notice that Eq.\,(\ref{eq:exmom1}) converges only for $n> 1/2$, since the Riemann zeta function diverges when $2 \,n=1$. We make the assumption that the asymptotic moments given in Eq. (\ref{eq:exmom1}) correspond to the moments of a function $\eta_A(t)$ defined as 
\begin{equation}\label{eq:mean1}
    \langle t^n\rangle_{A} = \int_0^\infty t^n \eta_A(t)dt \,.
\end{equation}
Explicitly, using Eq. (\ref{eq:exmom1}),
\begin{equation}
    \langle t^n\rangle_{A} = 2\tau_L^{n-\frac{1}{2}}\sqrt{\tau_0}n!\pi^{-2n}(2^{2n}-1)\zeta(2n).
\end{equation}
The naive concept is that if we find the function $\eta_A(t)$, it will also characterize the probability density function $\eta(t)$. However, as we will demonstrate below, this assumption introduces a subtle issue. Here the subscript A stands for asymptotic, since the moments $\langle t^n\rangle_{A}$ are those given in Eq. (\ref{eq:exmom1}), which is valid for large systems.
\par Recall the definition of the Mellin transform $\phi(z)$ of a function $f(t)$ is defined as $\phi(z)=\int_0^{\infty} t^{z-1}f(t)dt$. The Mellin transform of an exponential function is given by
 \begin{equation}\label{eq:mellinexp}
    \mathcal{M}\{e^{-\alpha t}\}(z) = \alpha^{-z}\Gamma(z),
\end{equation}
where $\Gamma(z)$ denotes the gamma function. From Eqs. (\ref{eq:mean1}-\ref{eq:mellinexp}), where we identify $n\rightarrow z-1$,
\begin{equation}\label{etaa}
    \eta_A(t)= 2\sqrt{\tau_0}\tau_L^{-3/2}\sum_{m=0}^{\infty} \left(\pi(m+\frac{1}{2})\right)^{2}e^{-\left(\pi^2(m+\frac{1}{2})^2\right)t/\tau_L }\,.
\end{equation}
Summing Eq. (\ref{etaa}), we eventually obtain
\begin{equation}\label{etaaf}
    \eta_A(t)= -\sqrt{\frac{\tau_0}{\tau_L}}\partial_{t }\vartheta_2(e^{-\pi^2 t/\tau_L}) \,,
\end{equation}
It is easy to verify that Eq. (\ref{etaaf}) gives the moments $\langle t^n \rangle_A $ in Eq. (\ref{eq:exmom1}). We argue that the function $\mathcal{I}(\tau)$ defined in Eq.\,(\ref{eq:infden1d}) can be used in principle to find the large $L$ limit of the moments. Hence it follows that this function is related to $\eta_A(t)$, and using Eq. (\ref{etaaf}) we find the main result of this section
  \begin{eqnarray}\label{inf10}
  \mathcal{I}(\tau)= -\partial_{\tau }\vartheta_2(e^{-\pi^2 \tau})\, ,
\end{eqnarray}
as found in Eq.\,(\ref{eq:inf1first}) in sec.\,\ref{sec:1D-diff}.

\section{Particle in a box - direct calculation of the infinite density}\label{appendix:1d}
The FPT PDF is given by $\eta(t)=-\partial_t \int_0^L P(x,t)dx$, where $P(x,t)$ is the solution for Eq. (\ref{eq:FPeq1}) given in sec. \ref{sec:1D-diff}. Using the eigenfunction expansion, we obtain   
\begin{equation}\label{eta1}
    \eta(t)=\frac{2 D}{L^2}\sum_{n=0}^{\infty} \pi(n+\frac{1}{2}) \sin\left(\pi(n+\frac{1}{2})\frac{x_0}{L}\right) e^{-D \pi^2(n+\frac{1}{2})^2 t/L^2} \, .
\end{equation}
which we can rewrite as
 \begin{equation}\label{eq:2sums}
    \eta(t)=\frac{D}{L^2}\sum_{n=0}^{n_c} c_n e^{-D \pi^2(n+\frac{1}{2})^2 t/L^2}+\frac{D}{L^2}\sum_{n=n_c}^{\infty} c_n e^{-D \pi^2(n+\frac{1}{2})^2 t/L^2} \, .
\end{equation}
 where $c_n=2\pi(n+1/2) \sin[\pi(n+1/2)x_0/L]$. In Eq. (\ref{eq:2sums}), the initial condition $x_0$ only enters at $c_n$. Assuming that $L\gg x_0$,  we can expand Eq. (\ref{eq:2sums}) using $c_n\simeq 2\pi^2(n+1/2)^2x_0/L$. This expansion is only valid where $\pi(n+1/2)x_0/L < 1$, meaning for $n<n_c$ when $n_c \sim L/x_0$. For the second sum in Eq. (\ref{eq:2sums}), the largest term is given for $n = n_c$. For large times $t\sim L^2/D$, and $n\geq n_c$, the exponential is vanishingly small, hence all following terms in the sum are negligible. 
 Since we consider the limit of large system, we can take $n_c\rightarrow \infty$, so Eq. (\ref{eta1}) can be approximate as 
 \begin{equation}\label{eta11}
    \eta(t)\simeq \frac{2 Dx_0}{L^3}\sum_{n=0}^{\infty}  \pi^2(n+\frac{1}{2})^2 e^{-D \pi^2(n+\frac{1}{2})^2 t/L^2} \, .
\end{equation}
Eq. (\ref{eta11}) can also be written as
\begin{equation}\label{eq:1detatau}
   \eta(t)\simeq   -2\sqrt{\frac{\tau_0}{\tau_L}} \partial_{t}\sum_{n=0}^{\infty}  e^{- \pi^2 \left(n+\frac{1}{2}\right)^2 t/\tau_L} \, ,
\end{equation}
where $\tau_L=L^2/D$ and $\tau_0=x_0^2/D$. It is useful to define a random variable $\tau\equiv t/\tau_L$ as the rescaled time, so in the limit of $L\rightarrow\infty$ and $t\rightarrow \infty$, this new variable remain finite. In this limit we can define the infinite density function using the asymptotic of the first passage time PDF as 
\begin{equation}
\mathcal{I}(\tau)\equiv\lim_{t,\tau_L \rightarrow \infty} \tau_L^{3/2}\tau_0^{-1/2}\, \eta(t),
\end{equation}
where, as stated before, the ratio $t/\tau_L$ remains finite. Summing the series in Eq. (\ref{eq:1detatau}), we obtain
\begin{equation}\label{eq:inf1}
\mathcal{I}(\tau)= -\partial_{\tau }\vartheta_2(e^{-\pi^2 \tau}) \, ,
\end{equation}
where $\vartheta_2(.)$ is the Jacobi elliptic theta function. Here we obtained the results presented in Eq. (\ref{eq:inf1}).

\section{Moments of the FPT}\label{appendix:1dmom}

Given our exact expression for $\eta(t)$ in Eq.\,(\ref{eta1}), we can write an exact expression for the moments:
\begin{align}
\langle t^q \rangle &=\int_0^\infty dt\, t^q\, \sum_{n=0}^\infty  \frac{D\sin\left(\left(n+\frac{1}{2}\right)\frac{\pi x_0}{L}\right)(2n+1)\pi}{L^2} e^{-(n+1/2)^2\pi^2 Dt /L^2}  \\
&= \sum_{n=0}^\infty \left(\pi^2(n+1/2)^2 D/L^2\right)^{-q-1} \Gamma(q+1) \frac{D\sin\left(\left(n+\frac{1}{2}\right)\frac{\pi x_0}{L}\right)(2n+1)\pi}{L^2} \\
&= 2\Gamma(q+1) \pi^{-1-2q}  (L^2/D)^{q}\sum_{n=0}^\infty \frac{\sin((n+1/2)\pi x_0/L)}{(n+1/2)^{2q+1}} 
\end{align}
Again, when $q>1/2$, the sum without the {\it sin} converges, and thus we can expand the {\it sin}, yielding exactly what we calculated above.
For $q<1/2$, we can approximate the sum by an integral over $n$ from 0 to $\infty$, also yielding our previous result.

We can do better by breaking the sum at $n=N$ where $1\ll N \ll L/x_0$.  We treat $q<1/2$ first. For the sum up to $N$ we can expand the {\it sin} giving
\begin{align}
S_1 &= \sum_{n=0}^N \frac{\sin((n+1/2)\pi x_0/L)}{(n+1/2)^{2q+1}}
 \\
&\sim \pi \frac{x_0}{L} \sum_{n=0}^N \frac{1}{(n+1/2)^{2q} }
\\
&\sim \pi \frac{x_0}{L} \left[ \frac{N^{1-2q}}{1-2q} + \left(2^{2q}-1\right)\zeta(2q) \right]
\end{align}
The second part of the sum can be well approximated by an integral, defining $y\equiv x_0\pi/L$:
\begin{align}
S_2 &\sim \int_N^\infty dk \frac{\sin(ky)}{k^{2q+1}}\\
&= \frac{\sin Ny}{2q N^{2q} } + y \int_N^\infty dk \frac{\cos ky}{2q k^{2q}} \\
&\sim \frac{y}{2q} N^{1-2q} + \frac{y}{2q}\int_0^\infty dk \frac{\cos ky}{ k^{2q}} - \frac{y}{2q} \int_0^N dk \frac{\cos ky}{ k^{2q}}\\
&\sim \frac{y}{2q} N^{1-2q} + \Gamma(1-2q) \frac{y^{ 2 q}}{2q} \sin(\pi q) - y\frac{N^{1-2q}}{(1-2q)}\\
&\sim  -\Gamma(-2q) y^{ 2 q} \sin(\pi q) - y\frac{N^{1-2q}}{1-2q}
\end{align}
Putting this together, we have
\begin{equation}
\langle t^q \rangle \sim \left[ 2\Gamma(q+1) \pi^{-1-2q}  (L^2/D)^{q}\right] \left[ -\Gamma(-2q) y^{ 2 q} \sin(\pi q)  + y \left(2^{2q}-1\right)\zeta(2q)\right]
\end{equation}
The first term exactly reproduces our previous result and the second is a correction, down by a factor of $(x_0/L)^{1-2q}$.  In particular, as $q\to 1/2$, the two terms approach the same magnitude and in fact cancel to leading order, leaving a $\ln L/x_0$ remainder. Taking the limit we get
\begin{equation}
\langle t^{1/2}\rangle \sim \sqrt{L^2/D} \pi x_0/L (1 + \ln(4L/x_0))/\pi^{3/2}
\end{equation}

For $q>1/2$, we can do the same thing.  For $S_1$, we have
\begin{align}
S_1 &\sim \pi \frac{x_0}{L} \sum_{n=0}^N \frac{1}{(n+1/2)^{2q} } \\
&\sim \pi \frac{x_0}{L} \left[(2^{2q}-1) \zeta(2q) - \frac{N^{1-2q}}{2q-1}\right]
\end{align}
The first term is independent of $N$ in the large-$N$ limit, and just reproduces our original $q>1/2$ calculation.  For $S_2$ and $q<1$, we have
\begin{align}
S_2  &\sim \frac{\sin Ny}{2q N^{2q} } + y \int_N^\infty dk \frac{\cos ky}{2q k^{2q}} \\
&= \frac{y}{2q} N^{1-2q} + \frac{y}{2q} \frac{\cos Ny}{(2q-1)N^{2q-1}} - \frac{y^2}{2q(2q-1)}\left[\int_0^\infty dk\, \frac{\sin ky}{k^{2q-1}} -
\int_0^N dk\, \frac{\sin ky}{k^{2q-1}}\right] \\
&\sim \frac{y}{2q} N^{1-2q} + \frac{y}{2q} \frac{1}{(2q-1)N^{2q-1}} - \frac{y^2}{2q(2q-1)} \left[ y^{2q-2} \Gamma(2 - 2 q) \sin(\pi q) - yN^{3-2q}\right]
&\sim \frac{y}{2q-1} N^{1-2q} - y^{2q}\Gamma(2-2q)\sin(\pi q)
\end{align}
Adding the two terms, we get that
\begin{equation}
\langle t^q \rangle \sim \-  2\Gamma(q+1) \pi^{-1-2q}  (L^2/D)^{q}\left[\pi \frac{x_0}{L} \left(2^{2q}-1\right) \zeta(2q) - (x_0/L)^{2q}\Gamma(2-2q)\sin(\pi q)\right]
\end{equation}

\section{Asymptotics of the moments amplitude for $q\rightarrow \infty$}\label{appendix:momamp}

We study the behavior of the moment amplitude \( M_q \) for large \( q \) in a general model. For \( \mathcal{I}(\tau) \) of the form
\begin{equation}
\mathcal{I}(\tau) = \sum_{n} c_n e^{-\lambda_n \tau},
\end{equation}
where \( {\lambda}_n \) are the eigenvalues associated with the system and \( c_n \) are the corresponding coefficients. For large \( q \), the moments are dominated by the longest times, which correspond to the smallest eigenvalue \( \lambda_0 \). Thus, we approximate the moment amplitude as  
\begin{equation}
\int_0^\infty d\tau \, \tau^q \mathcal{I}(\tau) \approx c_0\int_0^\infty d\tau \, \tau^q e^{-\lambda_0 \tau}.
\end{equation}

Applying Laplace's method, we rewrite the integrand as  
\begin{equation}
I_q = \int_0^\infty d\tau \, e^{-\lambda_0 \tau + q \log{\tau}},
\end{equation}
where the exponent is given by  
\begin{equation}
\Phi(\tau) = -\lambda_0 \tau + q \log{\tau}.
\end{equation}
The saddle point occurs at the maximum of \( \Phi(\tau) \), determined by setting its derivative to zero:  
\begin{equation}
\Phi'(\tau^*) = -\lambda_0 + \frac{q}{\tau^*} = 0,
\end{equation}
yielding  
\begin{equation}
\tau^* = \frac{q}{\lambda_0}.
\end{equation}
Expanding \( \Phi(\tau) \) around the saddle point gives  
\begin{equation}
\Phi(\tau) \approx \Phi(\tau^*) - \frac{\lambda_0}{2 q} (\tau - \tau^*)^2.
\end{equation}
The integral is then approximated as a Gaussian,  
\begin{equation}
I_q \approx e^{\Phi(\tau^*)} \int_{-\infty}^{\infty} d\tau \, e^{- \frac{\lambda_0}{2 q} (\tau - \tau^*)^2} = e^{\Phi(\tau^*)} \sqrt{\frac{2 \pi q}{\lambda_0}},
\end{equation}
where  
\begin{equation}
\Phi(\tau^*) = - q + q \log{\left( \frac{q}{\lambda_0} \right)}.
\end{equation}
Collecting the terms, we obtain the asymptotic form  
\begin{equation}
\int_0^\infty d\tau \, \tau^q \mathcal{I}(\tau) \sim e^{-q} q^q,
\end{equation}
which diverges as \( q \to \infty \).

\section{Two-dimensional wedge}\label{appendix:2dwedge}

The diffusion Eq. in plane polar coordinates is given by
\begin{equation}\label{eq:FPwedge}
    \frac{\partial}{\partial t}P(r,\phi,t)=D\left(\partial_r^2+\frac{1}{r}\partial_r+\frac{1}{r^2}\partial_{\phi}^2\right)P(r,\phi,t) \, ,
\end{equation}
where $\phi$ is the polar angle coordinate. We assume separation of variables to write the solution as $P(r,\phi,t)=R(r)\Phi(\phi)\exp(-D\lambda t)$, where $R(r)$ are the radius eigenfunctions and $\Phi(\phi)$ the angular eigenfunctions. Through the boundary conditions $\Phi(\phi=0)=\Phi(\phi=\pi/\nu)=0$, we reach that the angular eigenfunctions are sine waves, that is, $\Phi(\phi)\sim \sin(\phi m \nu)$ where $m$ is an integer.
By choosing the initial condition to be $P(r,\varphi,t=0)=\nu \sin(\phi \nu)\delta(r-r_0)/ 2 r_0 $, we can reduce Eq. (\ref{eq:FPwedge}) into an effective one-dimensional radial problem as only the $m=1$ term is not null. The general solution for the radial part is given by a combination of first and second kind Bessel functions. Since the domain includes $r=0$, where the second kind Bessel diverges, we will discard these solutions, with only the first kind term remaining. The solution is then given by
\begin{equation}
    P(r,\phi,t)=\sin(\nu \phi )\sum_k \frac{J_{\nu}(\frac{kr_0}{L})}{\int_0^L r J^2_{\nu}(\frac{kr}{L})dr }  J_{\nu}(\frac{kr}{L}) e^{-D \frac{k^2}{L^2} t} \,
\end{equation}
where $k$ is obtained by applying the reflective boundary, namely a  no-flux boundary condition at \(r = L\). Mathematically, this condition can be written as:
\[
\mathbf{J}_r = -D \frac{\partial P(r, \theta, t)}{\partial r} = 0 \quad \text{at} \quad r = L,
\]
where \(\mathbf{J}_r\) is the radial probability flux, thus, the reflecting boundary condition at \(r = L\) is:
\[
\frac{\partial J_\nu(k r/L)}{\partial r}_{r=L}=0
\]
From that, $k$ follows the equation
\begin{equation}
    J_{\nu+1}(k)-J_{\nu-1}(k)=0  \, .
\end{equation}

The FPT PDF is given by $\eta(t)=-\partial_t S(t)$, where $S(.)$ is the survival probability, defined as $S(t)= \int_0^{\alpha}d\phi \int_0^L rdr P(r,\phi,t)$, then 
\begin{equation}\label{eq:etawedge}
    \eta(t)=\tau_L^{-1}\sum_k c_k k^2 J_{\nu}\left(k\sqrt{\frac{\tau_0}{\tau_L}}\right) e^{- k^2\frac{t}{\tau_L} } \, ,
\end{equation}
where $\tau_L = L^2 / D$ is the exponential decay time scale, $\tau_0 = r_0^2 / D$ is the diffusive time scale, and the coefficient $c_k$ is given by
\begin{equation}
    c_k\equiv \frac{\int_0^L r J_{\nu}(\frac{k r}{L}) dr}{\int_0^L r J_{\nu}^2(\frac{k r}{L}) dr} \, 
\end{equation}
\begin{equation}
     =\frac{2^{1-\nu} k^{1+\nu} }{(2+\nu) \Gamma(1+\nu)} \frac{ \text{}_p F_q(1 +\frac{\nu}{2}, \{1 + \nu,2 + \frac{\nu}{2}\}, -\frac{k^2}{4})}{k J_{\nu}^2(k)+k J_{\nu-1}^2(k)- 2 \nu J_{\nu}(k)J_{\nu-1}(k)} \, ,
\end{equation}
where $_pF_q({a_1,...,a_p},{b_1,...,b_q};z) = \sum_k (a_1)_k,...,(a_p)_k / (b_1)_k,...,(b_q)_k z^k/k!$  is the generalized hypergeometric function and $(a)_k = \Gamma(a+k)/\Gamma(a)$ is Pochhammer symbol.
Since both integrals at Eq. (\ref{eq:akwedge1}) proportional to $\sim L^2$, $c_k$ does not depend on $L$.

\section{FPT for CTRW}\label{appendix:ctrw}

\subsection{General transformation of the first-passage moments}\label{appendix_transformfrac}

The q-moment of the first-passage PDF is defined as
\begin{equation}
    \langle t^q\rangle_{\alpha}=\int_0^\infty t^q \eta_\alpha(t)dt
\end{equation}
plug in $\eta_{\alpha}(t)=-\partial_t\int_{0}^{L}P_{\alpha}(x,t)dx$, where $P_{\alpha}(x,t)$ defined in Eq. (\ref{eq:p_ctrw}), then 
\begin{equation}
    \langle t^q\rangle=-\int_{0}^{L}dx\int_{0}^{\infty}ds P_{1}(x,s)\int_0^\infty dt  \text{ }t^q \partial_t n(s,t)
\end{equation}
Plug in $n(s,t)$ from Eq. (\ref{eq:ctrw_n}) and use integration b.p

\begin{equation}
    =\frac{q}{\alpha}(K_\alpha/D)^{1/\alpha}\int_{0}^{L}dx\int_{0}^{\infty}ds P_1(x,s)s^{-(1+1/\alpha)}\int_0^{\infty} dt \text{ } t^q  l_{\alpha}(\frac{K_\alpha^{1/\alpha}t}{(D s)^{1/\alpha}}) 
\end{equation}
change variables to $y=\frac{K_\alpha^{1/\alpha}t}{(D s)^{1/\alpha}}$ and solve the integration over x and y
\begin{equation}
    =\frac{q}{\alpha}\int_{0}^{L}dx\int_{0}^{\infty}ds P_{1}(x,s)s^{-1+q/\alpha}\int dy  y^q l_{\alpha}(y)
\end{equation}

\begin{equation}
    =\langle t^{q}\rangle _{Levy}\frac{q}{\alpha}\int_{0}^{L}dx\int_{0}^{\infty}ds P_{1}(x,s)s^{-1+q/\alpha}
\end{equation}
\begin{equation}
    =\langle t^{q}\rangle _{Levy}\int_{0}^{\infty}ds \eta_{1}(s)s^{q/\alpha}
\end{equation}
\begin{equation}\label{eq:ctrw_exmom}
   \langle t^q\rangle_{\alpha} =\langle t^{q/\alpha}\rangle _1 \langle t^{q}\rangle _{Levy}
\end{equation}
where 
\begin{equation}\label{eq:levymom}
\langle t^{q}\rangle _{Levy} = \frac{\Gamma[-q/\alpha]}{\alpha \Gamma[-q]}
\end{equation}
are the moments of the one-sided L\'evy distribution, and 
\begin{equation}
\langle t^{q/\alpha}\rangle _1  =\int_0^\infty t^{q/\alpha}\eta_1(t)dt
\end{equation}

are the $q/\alpha $ moment of the first passage PDF relate to the solution of the ordinary Fokker-Planck Eq. (\ref{eq:FPeq}).

\subsection{Short-time limit of the infinite density for CTRW}\label{appendix:coeffctrw}
In the limit of continuous n, by changing the sum with integration, namely $2n+1\rightarrow k$, Eq. (\ref{eq:infctrw}) yields 
\begin{equation}
\mathcal{I}_{\alpha}(\tau) \rightarrow \frac{1}{2\pi } \int_0^{\infty}dk k^2 \tau^{\alpha-1}E_{\alpha,\alpha}(-k^2 \tau^{\alpha}).
\end{equation}
Performing a change of variables $y\equiv k\tau^{\alpha/2}$
\begin{equation}\label{eq:ctrw_shorttime}
= \frac{1}{2\pi }\tau^{-1-\alpha/2}I ,
\end{equation}
where the coefficient $I$ is given by the following integral 
\begin{equation}\label{eq:apeni}
I=\int_0^{\infty} y^2E_{\alpha,\alpha}(-y^2)dy
\end{equation}
change variables $y^2=x^{\alpha},dy = \alpha x^{\alpha/2-1}dx/2$, Eq. (\ref{eq:apeni}) gives
\begin{equation}\label{eq:appen_d_a}
=\frac{\alpha}{2}\int_0^{\infty} x^{3\alpha/2-1}E_{\alpha,\alpha}(-x^\alpha)dx
\end{equation}

on the other hand 
\begin{equation}
\int_0^{\infty}\frac{e^{-ux}-1}{u^{1+\alpha/2}} du= x^{\alpha/2}\int_0^{\infty}\frac{e^{-z}-1}{z^{1+\alpha/2}} dz= x^{\alpha/2}\Gamma(-\alpha/2).
\end{equation}
Here we used change of variables $z=ux$.
Plugging in $x^{\alpha/2}$ into Eq. (\ref{eq:appen_d_a})
\begin{equation}
I=\frac{\alpha}{2\Gamma(-\alpha/2)}\int_0^{\infty}dx\int_0^{\infty}du \frac{e^{-ux}-1}{u^{1+\alpha/2}} x^{\alpha-1}E_{\alpha,\alpha}(-x^\alpha)
\end{equation}
Using the Laplace transform  $\mathcal{L}[x^{\alpha-1}E_{\alpha,\alpha}(-x^{\alpha})](u)=1/(u^{\alpha}+1)$
\begin{equation}
=\frac{\alpha}{2\Gamma(-\alpha/2)}\int_0^{\infty}du u^{-1-\alpha/2} (1-\frac{1}{u^{\alpha}+1}), \end{equation}
then
\begin{equation}
I=\frac{\alpha\pi}{\Gamma(1-\alpha/2)}. \end{equation}

\section{Fractal geometry}\label{appendix:fractal}
The solution for the diffusion Eq. (\ref{eq:diff_frac}) is given by 

 \begin{equation}\label{eq:FPeq}
    P(r,t)=r^{(d_w-d)/2}\sum_n c_n \left( J_{1-\nu}(2\sqrt{\frac{\lambda}{d_w^2 K}}r^{d_w/2})+J_{\nu-1}(2\sqrt{\frac{\lambda}{d_w^2 K}}r^{d_w/2}) \right)e^{-\lambda_n t}
\end{equation}
where $\nu=d_f/d_w$. To have non-zero current at the origin $j(r=0)\neq 0$, we keep only the first term.
Assume absorbing boundary at $r=0$, meaning $P(0,t)=0$, and reflecting boundary at $r=L$, meaning $j(r=L)=0$, therefore
\begin{equation}
    \partial_r P(r,t)|_{r=L}\propto J_{-\nu}(2\sqrt{\frac{\lambda}{d_w^2 K}}L^{d_w/2})=0 
\end{equation}
then
\begin{equation}
    \sqrt{\lambda} =\frac{d_w \sqrt{K}}{2L^{d_w/2}}z_{-\nu,n}
\end{equation}

apply initial condition 
$P(r,0)=\delta(r-r_0)r^{1-d}$, we obtain
\begin{equation}
   c_n =\frac{L^{-d_w}}{a_n} r_0^{(d_w-d)/2} J_{1-\nu}\left(z_{-\nu,n}(\frac{r_0}{L})^{d_w/2}\right)
\end{equation}

where
\begin{equation}
   a_n=\int_0^1 dx x^{d_w-1} J_{1-\nu}^2\left(z_{-\nu,n}x^{d_w/2}\right)=\frac{2}{w}\int_0^1 dx x J_{1-\nu}^2\left(z_{-\nu,n}x\right)=\frac{1}{w}J_{1-\nu}^2(z_{-\nu,n}). 
\end{equation}

the FPT PDF defined as $\eta(t)=-\partial_t\int_0^L dr r^{d-1} P(r,t)$, is given now as
\begin{equation}
    \eta(t) = L^{(w+d)/2}\sum_n \lambda_n c_n b_n e^{-\lambda_n t}
\end{equation}

where
\begin{equation}
    b_n=\int_0^1 dx  x^{(w+d-2)/2} J_{1-\nu}(z_{-\nu,n}(x)^{d_w/2})=\frac{2^{1+\nu}}{d_w\Gamma(1-\nu)}z_{-\nu,n}^{-1-\nu}
\end{equation}

The PDF is then given by
\begin{equation}
    \eta(t) = L^{(d-3d_w)/2}r_0^{(d_w-d)/2}\frac{2^{\nu-1}K d_w^2}{\Gamma(1-\nu)} \sum_n  \frac{J_{1-\nu}\left(z_{-\nu,n}(\frac{r_0}{L})^{d_w/2}\right) }{J_{1-\nu}^2(z_{-\nu,n})} 
 z_{-\nu,n}^{1-\nu} e^{-\lambda_n t}
\end{equation}
which give the solution obtained by Meyer et al \cite{meyer2011universality} in the large domain limit $L\gg r_0$.

We can approximate the solution for short times by taking the limit of the higher eigenvalues to obtain the solution for an infinite system. At that limit, we can approximate the sum into integration and obtain
\begin{equation}
    \eta(t)\rightarrow \frac{K d_w^2 }{ \Gamma(\nu) \Gamma(2-\nu)}L^{d-2d_w}r_0^{d_w-d}  \int_0^{\infty} dk   k^{3-2\nu}  e^{-d_w^2 K \lambda t/(4L^{d_w})}
\end{equation}
\begin{equation}
    = \frac{K d_w^2 }{ \Gamma(\nu) }r_0^{d_w-d} ( d_w^2 K t)^{-2+\nu} /2   
\end{equation}
when the solution for an inifintie system is know to be \cite{fa2003power}
\begin{equation}
    \eta(t)_{\inf} =e^{-\frac{r_0^{d_w}}{Kd_w^2t}} \frac{K d_w^2 }{ \Gamma(\nu) }r_0^{d_w-d} ( d_w^2 K t)^{-2+\nu}/2  
\end{equation}

\section{Simulation on the Sierpinski gasket}\label{appendix:simulation}
The master Eq. for the probability $P_n(i)$ for a random walker on a graph $G$ to be at site $i$ at time $n$ is given in general by
   \begin{equation}
       P_{n+1}(i)=MP_{n}(i)
    \end{equation}
    where M is the adjacency matrix. Apply an absorbing boundary condition at the target site $i=i_{target}$, namely
     \begin{equation}
       P_{n}(i_{target})=0.
    \end{equation}
    The survival probability, namely the probability that the particle did not arrive to the target by the $n$-th jump is given by
    \begin{equation}\label{eq:survsirp}
        S_n=\sum_{i} P_n(i),
    \end{equation}
    then the FPT probability distribution is defined by the numeric derivative of Eq. (\ref{eq:survsirp}) as
 \begin{equation}
        \eta_n=\frac{S_{n}-S_{n-1} }{\Delta n}, 
    \end{equation}
    where $\Delta n=1$.

\section{Force field }\label{appendix:force}

\subsection{Harmonic potential}\label{appendix:HO}
The Fokker Planck operator for a particle under a harmonic potential, $F(x)=-k_B T \kappa x$, is given by 
\begin{equation}\label{eq:FPophar}
    \partial_t P(x,t)=D\partial_x(\kappa x+\partial_x) P(x,t)  \, .
\end{equation}
We consider the solution of Eq. (\ref{eq:FPophar}) in the interval $x\in[0,\infty)$, with the initial condition $P(x,0)=\delta(x-x_0)$ where $x_0>0$. Applying an absorbing boundary condition on the origin demand $P(x,t)|_{x=0}=0$. The solution may be written in terms of an eigenfunction expansion as $P(x,t)=\sum_k c_k \phi_k(x) e^{-D \lambda_k t}$, where $\phi_k(x)$ are the eigenfunctions solve the equation $\mathcal{L}_{fp}\phi_k=\lambda_k \phi_k$, with the corresponding set of eigenvalues $\lambda_k$. Since the FP operator in Eq. (\ref{eq:FPophar}) yields a second-order equation, it requires two linearly independent solutions, given by
\begin{equation}
    \phi_k(x)=e^{-\frac{\kappa x^2}{2 }}(\alpha H_{\frac{\lambda}{\kappa}}(\sqrt{\frac{\kappa}{2}}x)+\beta\, _1F_1(-\frac{\lambda}{2\kappa},\frac{1}{2},\frac{\kappa x^2}{2}) ) \, ,
\end{equation}
where $H_{\lambda/\kappa}(.)$ are the Hermite polynomials and $_1F_1(a,b,z)$ is the Kummer confluent hypergeometric function. The particle is subjected to an absorbing boundary at the origin, so we only consider eigenfunctions that follow $\phi_k(x)|_{x=0}=0$. This leaves us with only the odd solutions of the Hermite polynomials, so $\beta=0$ , $\alpha=1$ for the odd (integers) values of $\lambda/\kappa$, then the eigenvalues are $\lambda_n=(2n-1)\kappa$. The particle starts at time $t=0$ in the position $x=x_0$, close to the origin. The coefficients $c_k$ are determined by the initial condition as $\int_0^{\infty} dx \psi_k(x)P(x,0) $, where $\psi_k(x)$ are the solutions of the adjoint Fokker-Planck equation, which are $\psi_k(x) = \exp(\kappa x^2 /2 )\phi_k(x)$, then the solution for the FPE \ref{eq:FPophar} is given by
\begin{equation}\label{eq:harprob}
    P(x,t)=\sqrt{\frac{\kappa}{2\pi}}\sum_{n=0}^\infty \frac{H_{2n+1}(\sqrt{\frac{\kappa}{2}}x_0)}{ 2^{2n+1}(2n+1)! } H_{2n+1}(\sqrt{\frac{\kappa}{2}}x)e^{-\frac{\kappa x^2}{2}}e^{-D(2n+1)\kappa t} \, .
\end{equation}
The first passage PDF is given by $\eta(t)=-\partial_t S(t)=-\int_0^{\infty}dx D\partial_x(\kappa x+\partial_x) P(x,t)=-D\partial_x P(x,t)|_{x=0}$. Plugging Eq. (\ref{eq:harprob}) in $S(t)$, we obtain
\begin{equation}\label{eq:etahar}
    \eta(t)= \sum_{n=0}^\infty \frac{\tau_L^{-1}}{\Gamma(\frac{1}{2}-n)\Gamma(2n+1)}  H_{2n+1}(\sqrt{\frac{\tau_0}{2\tau_L}})e^{-(2n+1) t/\tau_L}\, .
\end{equation}
where $\tau_0 = x_0^2/D$ is the diffusive time scale as defined in section \ref{sec:1D-diff}, and $\tau_L = (D \kappa)^{-1}$ corresponds to the time scale dominant the long time behavior of the FP PDF, meaning that in the limit of $t\gg \tau_L$, most of the particles will be reflected back to the origin due to the potential. 

The moments of the first passage PDF, defined as $ \langle t^q\rangle=\int_0^\infty t^q \eta(t) dt$, when $q>0$, are calculated exactly from Eq. (\ref{eq:etahar}) as
\begin{equation}\label{eq:harmomex}
     \langle t^q\rangle=\Gamma(1+q)\tau_L^q\sum_{n=0}^\infty \frac{(2n+1)^{-1-q}}{\Gamma(1/2-n)\Gamma(2n+1)}  H_{2n+1}(\sqrt{\frac{\tau_0}{2\tau_L}}) \, .
\end{equation}

\section{Asymptotics of the moments for general force field around $q\rightarrow 1/2$}\label{appendix:wkbmom}

From Eq.\,(\ref{eq120}), the PDF for a general confining force field in the WKB approximation is given by
\begin{equation}
\eta(t) \sim \frac{x_0}{\pi L^3} \int_{\tilde{E}_0}^{\infty} d\Tilde{E} \sqrt{\Tilde{E}} e^{-\Tilde{E} t/L^2}\, .
\end{equation}
From that, the FPT moments are defined as
\begin{equation}
\langle t^q \rangle \sim \frac{x_0}{\pi L^3} \int_{\tau_0}^\infty dt \, t^q \int_{\tilde{E}_0}^{\infty} d\Tilde{E} \sqrt{\Tilde{E}} e^{-\Tilde{E} t/L^2}\, .
\end{equation}
Integrating over \(\tilde{E}\), we obtain
\begin{equation}
\langle t^q \rangle \sim \frac{x_0}{\pi L^3} \int_{\tau_0}^\infty dt \, t^q \left( \frac{\sqrt{\tilde{E_0}e^{-\tilde{E_0}\tau}}}{\tau}+\frac{\sqrt{\pi}Erfc(\sqrt{\tilde{E_0}\tau})}{2\tau^{3/2}} \right)
\end{equation}

Changing the variable to \(\tau = t/L^2\), we rewrite the expression as:

\begin{equation}
\langle t^q \rangle \sim \frac{x_0}{\pi L^3} L^{2q+2} \int_{\tau_0/L^2}^\infty d\tau \, \tau^q \left( \frac{\sqrt{\tilde{E}_0 e^{-\tilde{E}_0 \tau}}}{\tau} + \frac{\sqrt{\pi} \text{Erfc}(\sqrt{\tilde{E}_0 \tau})}{2\tau^{3/2}} \right).
\end{equation}

This simplifies to:

\begin{equation}
\langle t^q \rangle \sim \frac{x_0}{\pi} L^{2q-1} \left[\frac{\sqrt{\pi} (\tau_0/\tau_L)^{q-1/2} \text{Erfc}(\sqrt{\tilde{E}_0 \tau_0/L^2}) - 2q \tilde{E}_0^{1/2-q} \Gamma(q, \tilde{E}_0 \tau_0/L^2)}{1-2q}\right],
\end{equation}
where \(\tilde{E}_0 = L^2 E_0\) and \(\tau_L = L^2/D\).

In the high-temperature limit where \(\tilde{E}_0 \propto 1\), the expression becomes:

\begin{equation}
\langle t^q \rangle \sim \frac{x_0}{\pi} L^{2q-1} \left[\sqrt{\pi} \left(\frac{\tau_0}{\tau_L}\right)^{q-1/2} \text{Erfc}\left(\sqrt{\frac{\tilde{E}_0 \tau_0}{\tau_L}}\right) - 2q \tilde{E}_0^{1/2-q} \Gamma(q, \tilde{E}_0 \tau_0/\tau_L)\right].
\end{equation}

Further simplifying, we get:

\begin{equation}
\langle t^q \rangle \sim \frac{x_0 \tau_L^{q-1/2}}{\pi} \left[\frac{\sqrt{\pi/4}\,(\tau_0/\tau_L)^{q-1/2} \text{Erfc}(\sqrt{\tau_0/\tau_L}) - q \Gamma(q, \tau_0/\tau_L)}{1/2 - q}\right].
\end{equation}

For \(q > 1/2\), where \((\tau_0/\tau_L)^{q-1/2} \rightarrow 0\), the moments simplify to:

\begin{equation}
\langle t^q \rangle \sim \frac{x_0 \tau_L^{q-1/2}}{\pi} \frac{q \Gamma(q, \tau_0/\tau_L)}{q - 1/2},
\end{equation}
which in the limit \(\tau_0/\tau_L \rightarrow 0\) becomes:

\begin{equation}
\sim \frac{x_0 L^{2q-1}}{\pi} \frac{\Gamma(1+q)}{q - 1/2}.
\end{equation}

For \(q < 1/2\), the moments are given by:

\begin{equation}
\langle t^q \rangle \sim \frac{x_0 \tau_L^{q-1/2}}{\pi} \frac{\sqrt{\pi/4}\,(\tau_0/\tau_L)^{q-1/2} \text{Erfc}(\sqrt{\tau_0/\tau_L})}{1/2 - q}.
\end{equation}

Thus, we have:

\begin{equation}
\langle t^q \rangle \sim \frac{x_0^{2q}}{2\sqrt{\pi}} \frac{1}{1/2 - q}.
\end{equation}

To conclude, for $q\rightarrow 1/2$, the moments are given by 
\begin{equation}
\langle t^q \rangle \propto  |q-1/2|^{-1} \begin{cases}
                    x_0^{2q} , &  q<1/2\\
                      x_0\,L^{2q-1} &  q>1/2\\
                    \end{cases}
\end{equation}


\begin{thebibliography}{10}
\bibitem{wedgenote}
The parameters $k$ and $c_k$ were obtained straightforwardly using the python
  scipy library.
\end{thebibliography}
\end{document}